\documentclass[prx,twocolumn,aps,epsf,showpacs,superscriptaddress,longbibliography,footinbib]{revtex4-1}
\usepackage[pdftex]{graphicx}
\usepackage{titlesec}
\titleformat{\paragraph}[runin]
        {\bfseries}
        {}
        {0.0em}
        {}
        [ --~~]
\titlespacing*{\paragraph}{0pt}{4pt}{0pt}

\usepackage{overpic}
\usepackage{dcolumn}
\usepackage{epsfig}
\usepackage{latexsym} 
\usepackage{amsmath,amssymb}
\usepackage{color}
\usepackage{array}
\usepackage{wasysym}
\usepackage{bbm} 
\usepackage{hyperref}
\usepackage{color}
\usepackage{array}
\usepackage{xcolor}
\usepackage{comment}
\usepackage{setspace}

\usepackage{booktabs}
\newcolumntype{C}[1]{>{\centering\arraybackslash}m{#1}}
\AtBeginDocument{
\heavyrulewidth=.08em
\lightrulewidth=.05em
\cmidrulewidth=.03em
\belowrulesep=.65ex
\belowbottomsep=0pt
\aboverulesep=.4ex
\abovetopsep=0pt
\cmidrulesep=\doublerulesep
\cmidrulekern=.5em
\defaultaddspace=.5em
}
\usepackage{booktabs}
\newcolumntype{R}[1]{>{\raggedleft\arraybackslash}p{#1}}
\AtBeginDocument{
\heavyrulewidth=.08em
\lightrulewidth=.05em
\cmidrulewidth=.03em
\belowrulesep=.65ex
\belowbottomsep=0pt
\aboverulesep=.4ex
\abovetopsep=0pt
\cmidrulesep=\doublerulesep
\cmidrulekern=.5em
\defaultaddspace=.5em
}

\newcommand{\be}{\begin{eqnarray}}
\newcommand{\ee}{\end{eqnarray}}

\newcommand{\<}{\langle}

\renewcommand{\>}{\rangle}
\renewcommand{\(}{\left(}
\renewcommand{\)}{\right)}
\renewcommand{\[}{\left[}
\renewcommand{\]}{\right]}
\renewcommand{\v}[1]{\boldsymbol{#1}} 

\newcommand{\bs}[1]{\boldsymbol{#1}}
\renewcommand{\d}{\partial}

\newcommand{\eps}{\epsilon}

\newcommand{\Z}{\mathbb{Z}}

\newcommand{\T}{\mathcal{T}}

\newcommand{\ident}[0]{\mathbbm{1}}

\newcommand{\com}[2]{\left[ #1, #2 \right]}

\newcommand{\abs}[1]{\left| #1 \right|}

\newcommand{\vpd}{\vphantom{\dagger}}
\newcommand{\micromo}{Q}

\newcommand{\InitRot}{W}
\newcommand{\Dop}{D}

\newcommand{\vecsym}[1]{\boldsymbol{#1}}

\newcommand{\vecth}{\vecsym{\theta}}

\newcommand{\vecomega}{\vecsym{\omega}}
\renewcommand{\vec}[1]{\bs{#1}}
\newcommand{\vecn}{\vecsym{n}}
\newcommand{\vecm}{\vecsym{m}}

\newcommand{\uvec}{\hat{\v{e}}}

\usepackage{xcolor}

\begin{document}
\title{Topological edge modes without symmetry in quasiperiodically driven spin chains}
	\author{Aaron J. Friedman }
	\affiliation{Department of Physics, University of Texas at Austin, Austin, TX 78712, USA}
	\author{Brayden Ware}
	\affiliation{Department of Physics, University of Massachusetts, Amherst, Massachusetts 01003, USA}
	\author{Romain Vasseur}
	\affiliation{Department of Physics, University of Massachusetts, Amherst, Massachusetts 01003, USA}
	\author{Andrew C. Potter}
	\affiliation{Department of Physics, University of Texas at Austin, Austin, TX 78712, USA}
\begin{abstract}
We construct an example of a 1$d$ quasiperiodically driven spin chain 
whose edge states can coherently store quantum information, protected by a combination of localization, dynamics, and topology. Unlike analogous behavior in static and periodically driven (Floquet) spin chains, this model does not rely upon microscopic symmetry protection: Instead, the edge states are protected purely by \emph{emergent} dynamical symmetries. We explore the dynamical signatures of this Emergent Dynamical Symmetry-Protected Topological (EDSPT) order through exact numerics, time evolving block decimation, and analytic high-frequency expansion, finding evidence that the EDSPT is a stable dynamical phase protected by bulk many-body localization up to (at least) stretched-exponentially long time scales, and possibly beyond. We argue that EDSPTs are special to the quasiperiodically driven setting, and cannot arise in Floquet systems. Moreover, we find evidence of a new type of boundary criticality, in which the edge spin dynamics transition from quasiperiodic to chaotic, leading to bulk thermalization.
\end{abstract}
\maketitle

Edge states of $1d$ topological phases can coherently store quantum information in a manner that is protected against stray fields, uncontrolled interactions, and crosstalk, making them promising candidates for quantum memory. In isolated and many-body localized (MBL) systems~\cite{DavidRahulReview,Vasseur_2016,GOPALAKRISHNAN20201,RevModPhys.91.021001}, this protection can extend to highly excited states~\cite{PhysRevB.88.014206,Bauer_2013,Bahri:2015aa,PhysRevB.89.144201,potter2015protection,2015arXiv150505147S,khemani2016phase}, enabling topological quantum memories without the need for cooling or ground state preparation. 
However, both fundamental and practical considerations restrict MBL protection to \emph{bosonic} systems, which, for $1d$ time-independent and Floquet systems, only admit a weaker form of symmetry-protected topological (SPT) order. Namely: (\emph{i}) symmetry restrictions on MBL preclude realizing fermion topological phases~\cite{potter2016symmetry} and (\emph{ii}) the atomic, molecular, and optical (AMO) platforms capable of realizing the spatiotemporal control of interactions required to synthesize complex phases (like superconducting qubits and circuit QED systems~\cite{blais2020circuit}, trapped ions~\cite{bruzewicz2019trapped}, and Rydberg atoms~\cite{saffman2010quantum,browaeys2020many}) are all comprised of bosonic degrees of freedom (qubits, spins, or oscillators). 

The prototypical example of an MBL-protected SPT phase~\cite{PhysRevB.83.035107,doi:10.1146/annurev-conmatphys-031214-014740,2013arXiv1301.0330T} is the AKLT (aka ``cluster state'' or Haldane phase) model~\cite{PhysRevLett.50.1153,PhysRevLett.59.799}, whose projective (``spin half") edge states are protected by two $\Z_2$ spin-rotation symmetries; this $\Z_2 \times \Z_2$ symmetry forbids any symmetric coupling from dephasing or depolarizing the edge states. However, this complicated symmetry is physically unnatural in most AMO systems: achieving it would require fine tuning, leaving the edge states vulnerable to many perturbations.

Time-periodic (Floquet) driving can actually simplify the symmetry requirements. In the analogous, dynamical Floquet SPT (FSPT), one of the microscopic $\Z_2$ symmetries is replaced by an \emph{emergent} symmetry arising from the drive's discrete time translation invariance. This emergent dynamical symmetry cannot be broken by local, time-periodic perturbations. Formally, FSPT phases are classified by extending the microscopic symmetry group to include time translation symmetries, i.e., the group, $\Z$, of translations by integer multiples of drive period, $T$~\cite{von2016phase,else2016classification,potter2016classification,roy2016abelian}. Physically, the Floquet cluster state undergoes a repeating topological spin echo process that dynamically decouples the edge spins from all perturbations that respect the microscopic $\Z_2$ symmetry~\cite{kumar2018string}. Crucially, unlike an ordinary spin echo sequence, interactions collectively stabilize the topological edge state motion against generic symmetric perturbations.

This construction begs the question: Can one forego microscopic symmetries entirely and engineer ``absolutely stable"~\cite{von2016absolute} dynamical topological phases protected only by \emph{emergent} dynamical symmetries? To this end, we consider generalizing periodic (single-tone) drives to \emph{quasi}periodic drives (comprising two tones with incommensurate periods; see, e.g.,~\cite{PhysRevX.7.041008, dumitrescu2018logarithmically, PhysRevB.98.220509,else2020long}). In analogy to spatial quasicrystals, the quasiperiodic drive can be viewed as a projection from a higher-dimensional ``time torus" with independent time translation directions for each drive tone. 
Previous work has shown that quasiperiodic driving enables new examples of dynamical symmetry breaking (e.g., time quasicrystals)~\cite{dumitrescu2018logarithmically} and SPT phases~\cite{else2020long}, but has so far overlooked the possibility of dynamical topological phases without microscopic symmetry protection.

Our strategy will be to replace the two $\Z_2$ symmetries that protect the static AKLT phase with emergent dynamical symmetries enforced by the drive.  We construct an explicit spin model with this property, and demonstrate the stability of edge states to generic quasiperiodic perturbations via numerical integration, time evolving block decimation (TEBD), and analytical methods. We refer to invertible (short-range entangled) dynamical topological phases protected solely by emergent dynamical symmetries (i.e. without any microscopic symmetries) as Emergent Dynamical Symmetry Protected Topological orders (EDSPTs). Interestingly, EDSPTs lie outside previously proposed formal classification schemes for (quasi)-periodically driven phases.

\begin{figure}[t]
\centering
\includegraphics[width = \columnwidth]{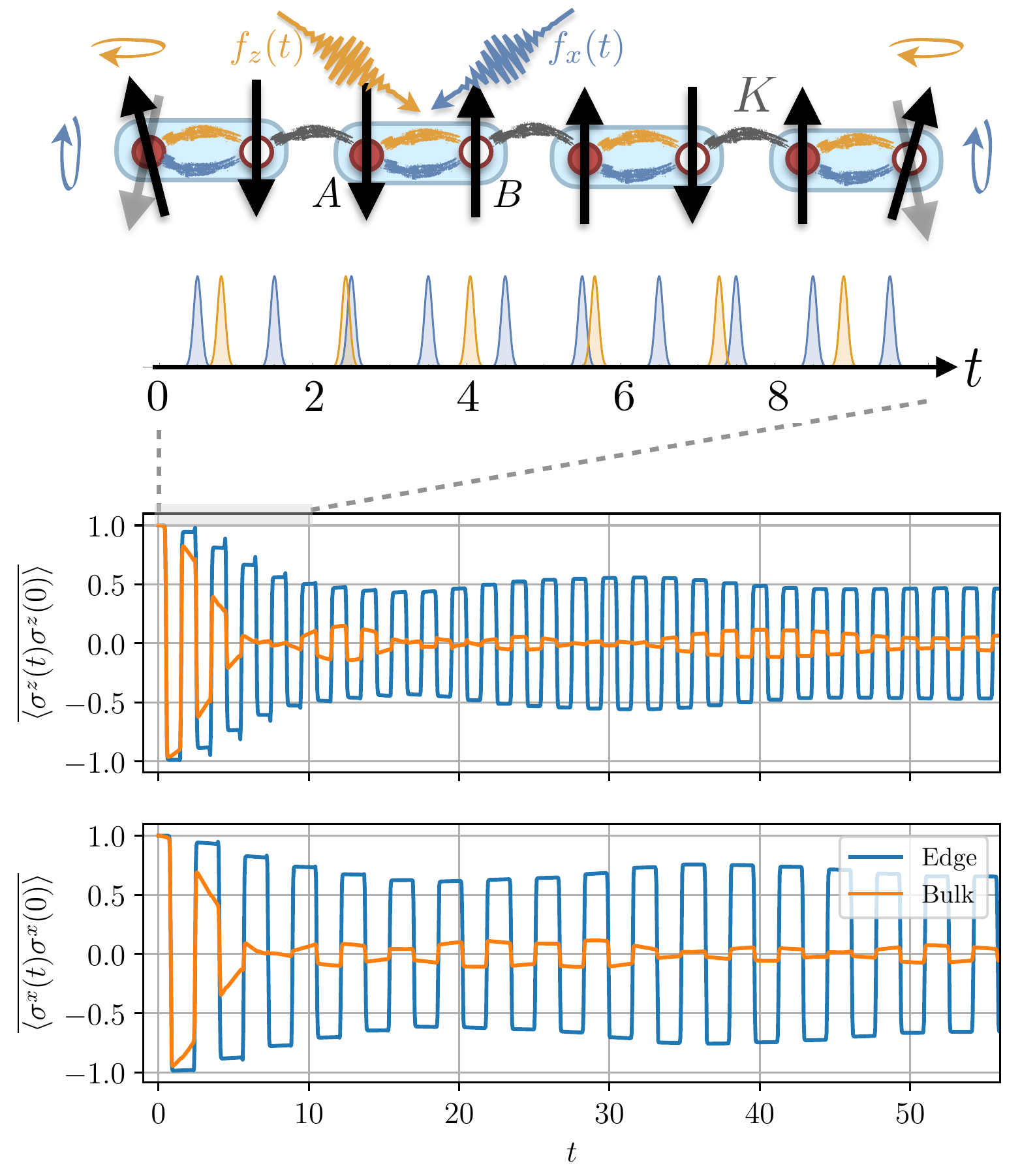} 
\caption{{\bf Ising EDSPT phenomenology---} 
\emph{Top}: Schematic of the EDSPT model in Eq.~\ref{eq:H}.
\emph{Bottom}: Time-evolving block decimation (TEBD) simulations of this model in a $50$-spin chain. Whereas the bulk spin correlators rapidly decay, edge spins exhibit long-lived, coherent, quasiperiodic oscillations indicative of their dynamical topological protection. Simulations use a time step $\Delta t = 0.002$, bond dimension $\chi\leq 1024$, truncation error $\eps \leq 10^{-8}$, and are averaged over $100$ disorder realizations; fo reach realization, the initial ($t=0$) state is an independently chosen, random $\sigma^z$-product state.
\label{fig:TEBD}
}
\end{figure}

We argue that EDSPTs are special to quasiperiodically driven systems. Analogous Floquet models can be continuously deformed to a trivial phase without a bulk phase transition, by applying a ``counter drive'' that neutralizes the edge dynamics. In the quasiperiodic setting, such counter drives lead instead appear to produce a bulk delocalization transition. Namely, we provide numerical evidence that attempting to cancel the topological edge dynamics of the quasiperiodic model necessarily results in strongly overlapping, noncommuting pulses that cause the edge dynamics to become chaotic, and subsequently heat up and melt the MBL bulk. This behavior appears unique to quasiperiodically driven systems, in which even a single spin can exhibit a $0d$ transition from quasiperiodic to chaotic motion~\cite{luck1988response,jauslin1991spectral,blekher1992floquet,jauslin1992generalized,crowley2019topological}.

\paragraph{Notion(s) of stability in quasiperiodic MBL}
Following~\cite{else2020long}, we define a $p$-tone quasiperiodic drive,
$H(t) = \sum_{\vec{n}\in \mathbb{Z}^p}e^{-i\vec{\omega}\cdot\vec{n}t} H^{\,}_{\vec{n}}$ (where the frequency vector $\vecomega$ has components $\omega^{\,}_i = 2\pi /T^{\,}_i$, $(i=1\dots p)$ with $\omega^{\,}_i / \omega^{\,}_j \notin \mathbb{Q}$), to be MBL if its time evolution, $U(t) = \mathcal{T} \, e^{-i \int_0^t H(t)dt}$, can be reduced to the form
\begin{align}
	U^{\,}_\text{Q-MBL}(t) = \mathcal{Q} (t) \, e^{-iD^{\,}_\text{MBL} t} \, \mathcal{Q}^\dagger (0),
	\label{eq:qmbl}
\end{align}
where the unitary, $\mathcal{Q} (t)$, is 
quasiperiodic in $t$, and $D^{\,}_\text{MBL}$ is a static MBL Hamiltonian with a complete set of local integrals of motion (LIOMs)~\cite{PhysRevLett.111.127201,PhysRevB.90.174202,PhysRevLett.117.027201}.

If $\mathcal{Q} (t)$ has the same quasiperiodicity as the drive Hamiltonian, $H(t)$, we say that the system preserves the dynamical symmetries.  Another possibility is a time quasicrystal with spontaneously broken dynamical symmetries~\cite{dumitrescu2018logarithmically}, wherein $\mathcal{Q} (t)$ is quasiperiodic but with an enlarged quasiperiodicity compared to $H(t)$. Like FSPTs, with closed boundary conditions, EDSPTs exhibit quasiperiodic MBL dynamics that preserve the dynamical symmetries; in open chains, the EDSPT's edge modes exhibit time quasicrystalline dynamics. Unlike static and Floquet evolution, quasiperiodically driving even a single spin can lead to chaotic dynamics~\cite{luck1988response,jauslin1991spectral,blekher1992floquet,jauslin1992generalized,crowley2019topological} that fail to reduce to the above form, instead realizing a continuous frequency spectrum. This possibility leads to an altered notion of stability for quasiperiodic MBL, since a single such chaotic spin can act as a continuous-spectrum noise source that can thermalize many otherwise-MBL spins.

In contrast to static systems~\cite{imbrie2016many}, there is no rigorous proof of stability of driven MBL phases. However, analytic arguments~\cite{ponte2015many,lazarides2015fate,ponte2015periodically} in favor of Floquet MBL (which can be supplemented by infinite-time numerical simulations), apply equally to smooth quasiperiodic drives (for which  reaching long times is numerically challenging). Recently established analytic bounds~\cite{else2020long,de2019very} (see also~\cite{PhysRevLett.115.256803,Abanin:2017aa,KUWAHARA201696} in the Floquet case) show that quasiperiodically driven disordered systems remain MBL at least up to stretched-exponentially long ``pre-heating" time scales, $\tau^{\,}_{\rm ph} \sim \exp(v^{-\gamma})$ (where $v$ is the appropriately normalized drive strength and $\gamma<1$ some exponent), and perhaps indefinitely. Throughout the following, we will assume that either quasiperiodically driven MBL is stable to infinitely long times, or that we are operating in a regime in which $\tau^{\,}_{\rm ph}$ significantly exceeds experimental time scales.

\paragraph{Model} 
Our starting point is an adaptation of the cluster state representation of the AKLT phase, defined on a spin chain with two sublattices ($A$ and $B$), each with $L$ spins:
\be 
	 H^{\,}_\text{CS} = -\sum\limits_{j=1}^{L-1}\sum\limits_{\mu=x,z}K^\mu_{j} \sigma^\mu_{B,j}\sigma^\mu_{A,j+1}
	 ~,~~~ 	
	 \label{eq:aklt} 
\ee
where $K^\mu_i$ are independently and identically distributed uniformly from $[\text{-}K,\text{-}K_\text{min}]\cup[K_\text{min},K]$~\footnote{To mitigate finite size effects in small scale numerical simulations, $|K_i|< K_\text{min}$ are excluded to avoid accidentally cutting the chain.}. Almost all eigenstates of $H^{\,}_\text{CS}$ exhibit an exact fourfold degeneracy corresponding to a pair of projective, zero energy (``spin half") edge spin operators: $\vec{\sigma}^{\,}_{A,1}$ and $\vec{\sigma}^{\,}_{B,L}$. The zero modes are protected by a pair of discrete $\Z_2$ spin-rotation symmetries generated by 
$g^{\,}_\mu = \prod_{j=1}^L \sigma_{A,j}^{\mu}\sigma_{B,j}^{\mu}$, for $\mu\in\{x,z\}$; together, these generate the symmetry group $\Z_2\times \Z_2$. The disordered couplings in Eq.~\ref{eq:aklt} ensure that the zero modes extend throughout a stable MBL phase in the presence of generic (time-independent) perturbations that respect this symmetry. 

To protect this phase dynamically (i.e., dispense with the microscopic symmetry requirements), 
we apply a quasiperiodic drive consisting of two tones with irrationally related periods, $T^{\,}_x = 1$, $T^{\,}_z = \varphi = \frac{1+\sqrt{5}}{2}$,
\begin{align}
	H^{\,}_0(t) &= \frac12\sum_{j=1}^L \(f^{\,}_x(t)\sigma^x_{A,j}\sigma^{x}_{B,j}+f^{\,}_z(t)\sigma^z_{A,j}\sigma^{z}_{B,j}\),
	\notag \\
	f^{\,}_\mu(t) &= \pi \sum_{n \in \Z} G^{\,}_w\(t-(n+\phi^{\,}_\mu)T^{\,}_\mu\) \label{eq:H0def}
\end{align}
where $G^{\,}_w(x) = \frac{1}{\sqrt{2\pi w^2}}e^{-x^2/2w^2}$ are normalized Gaussian pulses with width $w$, and $\vec{\phi}$ controls the phase of the drive (which takes values on the unit torus, $\mathbbm{T}^2$; unless otherwise specified, we choose $\phi^{\,}_x=\phi^{\,}_z=\frac12$). 
Since all of the terms in $H^{\,}_0$ commute, the resulting time evolution, $U^{\,}_0(t) = \T e^{-i\int_0^t H^{\,}_0(s)ds}$ is straightforward to compute. 

To motivate this construction, consider the single-tone (Floquet) limit by omitting the $f^{\,}_x$ drive, and taking $w\rightarrow0$ (i.e., $\delta$-function pulses). Here,  $H(t)=H^{\,}_0(t)+H^{\,}_\text{CS}$ realizes a Floquet SPT phase protected by a microscopic $g^{\,}_x$ symmetry. Each $z$-pulse in $H^{\,}_0$ has the same effect on the system as applying the symmetry generator $g^{\,}_z$. Suppose that we extend the model with a $g^{\,}_x$-preserving, $g_z$-breaking, perturbation, $V$, with $g^{\,}_zVg^{\,}_z=-V$. Roughly speaking, the net, $g^{\,}_z$-breaking effect of $V$ averages to zero after an even number of pulses, much like a spin echo pulse sequence. This cancellation effectively restores the $g^{\,}_z$-symmetry in a periodically rotating frame.

\begin{figure}[t]
\centering
\includegraphics[width = 1.0\columnwidth]{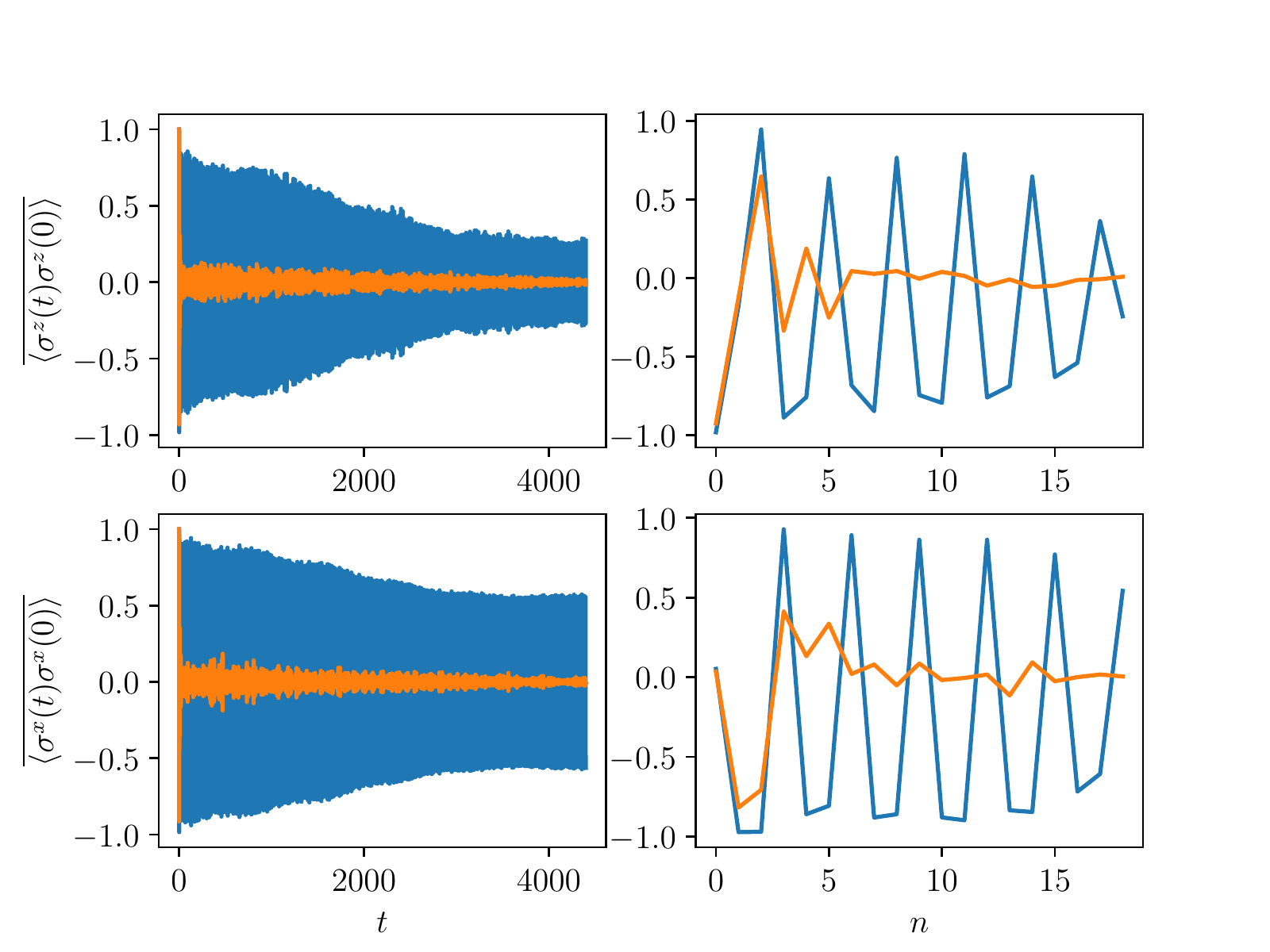} 
\caption{{\bf Topological edge response---} \emph{Left}: Numerical simulation of $\sigma^x$ and $\sigma^z$ correlations for $14$ spins with small, nonzero perturbations of all types, averaged over disorder and initial random $z$-basis product states. These simulations use Gaussian pulses with standard deviation of $5$\% of period, pulse area of $0.95 \pi$, $K=0.3$, and $J=B=0.05$. The bulk correlations decay rapidly, while the edge exhibits oscillations that saturate to $\sim 50\%$ amplitude and persist to the longest times simulated. \emph{Right}: Plot of the same correlations evaluated only at Fibonacci times; at successive Fibonacci times, $U^{\,}_0$ realizes a \emph{periodic} sequence of operators $\sigma^{x,y,z}$, resulting in periodic edge response at Fibonacci times.}
\label{fig:EDL14}
\end{figure}

Na\"ively, one can expect the two-tone drive (with both $f^{\,}_{x,z}$ pulse trains) to operate similarly, with both $x$, $z$ pulses spin echoing away perturbations that are odd under $g^{\,}_x$ or $g^{\,}_z$,  effectively imposing a dynamically enforced $\Z_2\times \Z_2$ symmetry. This conclusion is not entirely obvious, since the $x$ and $z$ pulses of the two-tone drive are quasiperiodically interleaved, such that $z$ pulses can come in between pairs of $x$ pulses, potentially interrupting the spin echo action. Despite this, we will show that the na\"ive argument above turns out to be essentially correct.

\paragraph{Solvable limit} Before addressing the generic stability of the model, we first examine a special, soluble limit that captures the characteristic phenomenology. The model $H^{\,}_0(t)+H^{\,}_\text{CS}$ is exactly soluble in the limit of infinitely thin $w\rightarrow 0$ pulses. The perfectly localized bulk is characterized by an extensive set of LIOMs~\cite{PhysRevLett.111.127201,PhysRevB.90.174202,PhysRevLett.117.027201}, $\sigma^\mu_{B,j}\sigma^\mu_{A,j+1}$, that commute with the pulses in $H^{\,}_0$. In contrast, the edge states, $\vec{\sigma}^{\,}_{A,1}$ and $\vec{\sigma}^{\,}_{B,L}$, are flipped about the $x$ and $z$ axes by the $f^{\,}_{x,z}$ pulses, respectively. This edge motion results in a quasiperiodic sequence of spin flips, which follow a Fibonacci sequence generated by
\begin{align} 
	\dots
	(\sigma^x\sigma^z\sigma^x\sigma^x\sigma^z)
	(\sigma^x\sigma^x\sigma^z)
	(\sigma^x\sigma^z)
	(\sigma^x)~,~~~
\end{align}
which includes the rotations due to all pulses up to some time $t$, where the parentheses show the Fibonacci recursion structure. The cumulative effect of these pulse sequences is to twirl the spin quasiperiodically about the $x,y,z$ axes, depending on the stopping time, $t$. This behavior will be crucial for stabilizing the edge motion when we consider perturbing the soluble limit considered here. 

We note that this quasiperiodic ``Pauli-twirling" pattern can be simply understood when examined at ``Fibonacci times," $t^{\,}_n \sim F^{\,}_n \approx \varphi F^{\,}_{n-1}$, defined as the times where $t$ and $t/\varphi$ are (locally) as close as possible to integers (the deviation from integer decays exponentially with $n$). For times $t=t^{\,}_n$, as a function of $n$, the edge spins are conjugated by a \emph{repeating} sequence of operators, $ \sigma^x\rightarrow \sigma^y \rightarrow \sigma^z \rightarrow \sigma^x  \cdots$, with periodicity three. This period tripling (in Fibonacci time) reflects the fact that an even number of $\pi$-pulses produce no effect, and that the Fibonacci numbers modulo two have a threefold periodic structure. In numerics, this edge-twirling pattern provides a convenient signature, similar to bulk signatures of time quasicrystals previously studied by two of us~\cite{dumitrescu2018logarithmically}.

\paragraph{Stability to generic perturbations}
To establish the stability of the idealized model, we now consider generic perturbations. The full model is given by
\begin{align}
	H(t) = H^{\,}_0(t) + H^{\,}_\text{CS} + V(t)
	\label{eq:H}
\end{align}
where $V(t)$ includes generic, local, quasiperiodic-in-time perturbations of strength $v\ll K$. We restrict to small, nonzero pulse widths $0<w\ll 1$ to limit the high frequency content of the drive.  While the analytic results presented are valid for any $V$ satisfying the above properties, for numerical simulation we specialize to
\begin{align}
	V(t) &= -\lambda H^{\,}_0(t) + \sum\limits_{j=1}^L \sum\limits_{\nu=x,z} J_j^\nu \sigma_{A,j}^\nu \sigma_{B,j}^\nu + \nonumber\\
	&~~~+ \sum\limits_{j=1}^L\sum\limits_{\alpha=A,B}\vec{h}^{\,}_{\alpha,j}\cdot\vec{\sigma}^{\,}_{\alpha,j}
\end{align}
where $\lambda$ is the deviation from perfect $\pi$ pulses, $J_j^\nu\sim [-J,J]$ terms compete with $H^{\,}_\text{CS}$ to give a nonzero correlation length, and the random fields, $h^{x,y,z}_{\alpha,j}\in [-h,h]$, break all microscopic symmetries.

Fig.~\ref{fig:TEBD} shows TEBD~\cite{PhysRevLett.91.147902,PhysRevLett.93.040502} simulations of large (50-spin, $L=25$) chains to moderate times ($t\sim 10^2$), and Fig.~\ref{fig:EDL14} shows exact numerical integration of time evolution for smaller (14-spin, $L=7$) chains to longer times ($t\sim 10^4$). To contrast the edge and bulk behavior, we consider two-point correlation functions $C_{\alpha,r}^{\mu}(t) = \overline{\<\sigma_{\alpha,r}^{\mu}(t)\sigma_{\alpha,r}^{\mu}(0)\>}$, where $\overline{(\dots)}$ denotes disorder and initial state averaging. The edge correlations initially decay before saturating to a nonzero value that persists up to the longest times simulated, indicating finite overlap with topologically protected edge states. In contrast, the (disorder averaged) bulk correlations quickly decay to zero due to sensitivity to local disorder couplings, signaling an absence of topological protection. Plotting the same data at Fibonacci times correctly accounts for the complicated quasiperiodic micromotion, revealing an underlying periodic oscillation due to the quasiperiodic twirling discussed above.

\paragraph{Topological edge state dynamics} To understand these results we employ the high-frequency (HF) expansion technique of~Ref.~\citenum{else2020long}, which allows the corresponding time evolution to be written (up to time $t\sim \exp\[(K/v)^{\gamma}\]$ with $\gamma \lesssim 2/3$ in our case~\cite{else2020long}, see also App. \ref{app:ElseDetails}) as
\begin{align}
	U (t) = \T \, \{e^{-i\int_0^tH(s)ds}\} = \InitRot^{\dagger} \, \micromo (t) \, U^{\,}_0 (t) \, e^{- i \Dop t} \, \InitRot ~~,~~ 
	\label{eq:hf}
\end{align}
where $W$ is a finite-depth local unitary, $Q(t)$ is a unitary with the same quasiperiodicity as $H^{\,}_0(t)$ satisfying $Q(0)=\mathbbm{1}$ (comparing to Eq.~\ref{eq:qmbl}, we have redefined $\mathcal{Q} (t)=\InitRot^{\dagger}\, Q(t) \, U^{\,}_0(t)$, such that $Q(t)$ has the same quasiperiodicity as $H(t)$). Explicit forms for $\Dop$, $Q$, $W$ can be computed order by order in $K,v$ (see App.~\ref{app:HFE} for details), and resemble those of more familiar Magnus expansions. 

For our model, $\Dop \approx H^{\,}_\text{CS} + \overline{V}^{\,}_\text{sym}$, results in a \emph{symmetrized} Hamiltonian that commutes with $g^{\,}_{x,z}$, and for $v\ll K$, takes the form of an MBL SPT Hamiltonian in the same phase as $H^{\,}_\text{CS}$ (weakly perturbed by local symmetric terms $\overline{V}_\text{sym}$ made from terms with typical norm $\sim v$). Thus, $U(t)$ is equivalent to an MBL SPT evolution in a quasiperiodically rotating frame, where the protecting symmetries are entirely emergent (i.e. may be completely broken by the Hamiltonian, $H(t)$, that generates $U(t)$). Specifically, the 
emergent protecting symmetry is generated by $\tilde{g}^{\,}_{x,z} = W \, g^{\,}_{x,z}\, W^\dagger$, i.e., ``locally dressed" versions of $g^{\,}_{x,z}$ with the same $\Z_2\times \Z_2$ group structure, whose precise form depends on $H(t)$. Crucially, weakly perturbing $H(t)$ by terms that do not commute with $g^{\,}_{x,z}$ merely modifies $W$ without undoing the existence or group structure of the emergent dynamical symmetry.

The primary signature and utility of this phase is its robust edge modes with topologically protected coherence. When $\Dop$ (Eq.~\ref{eq:hf}) lies in the SPT phase, it hosts edge modes, $\v{\Sigma}^{\,}_{L/R}$, that both transform projectively under the (emergent) $\Z_2\times \Z_2$ symmetry generated by $g^{\,}_{x,z}$, and are stable against \emph{any} quasiperiodic  perturbation to $H(t)$\footnote{That is to say, the perturbations need not commute with either $g^{\,}_{x,z}$, in contrast to the Floquet case.}. As with equilibrium SPTs, for $\overline{V}^{\,}_\text{sym}\neq 0$, the edge modes, $\v{\Sigma}_{L/R}$, 
are no longer simply the single-site operators $\sigma^{\,}_{A,1}, \sigma^{\,}_{B,L}$, but rather, get ``dressed'' with nearby operators, whose support decays exponentially with distance, $r$, into the bulk, as $e^{-r/\xi}$ (where $\xi$ is the localization length).

\begin{figure*}[tb]
\centering
\begin{overpic}[width=2.0\columnwidth]{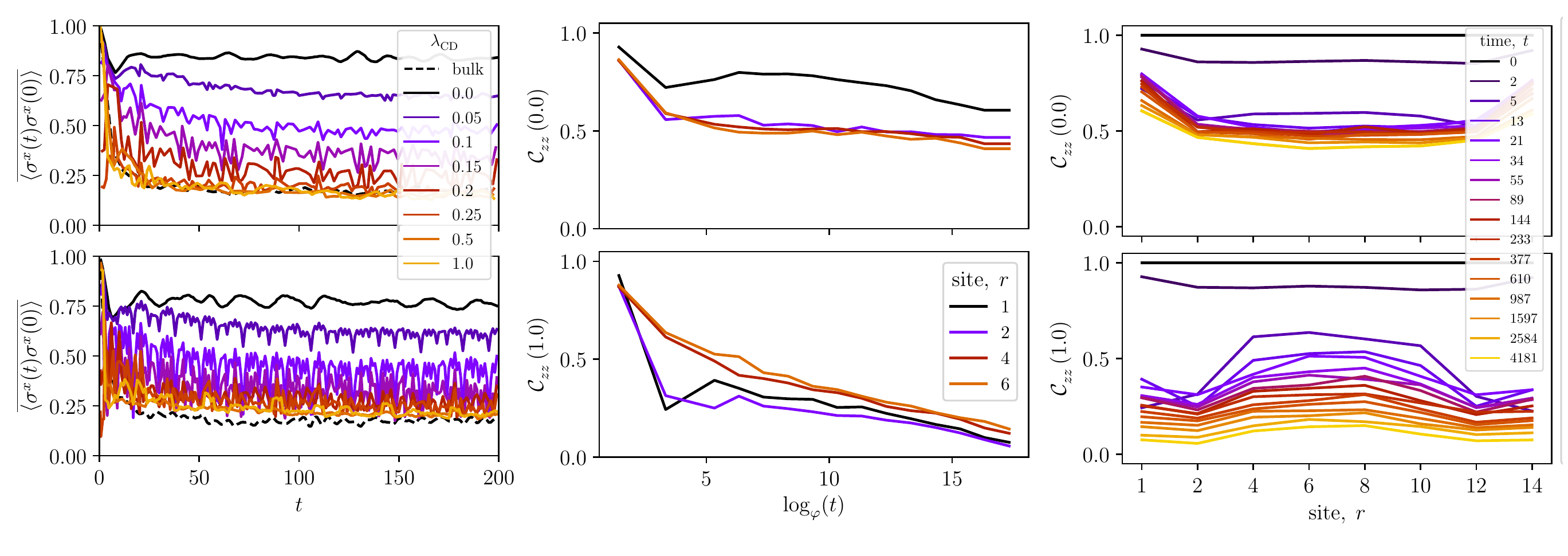}
\put (0,2) {(a)}
\put (34,2) {(b)}
\put (67,2) {(c)}
\end{overpic}
\caption{{\bf Boundary thermalization from counter driving ---}
(\emph{a}): Destruction of edge correlations for sufficiently strong counter drives. (\emph{b},\emph{c}) Long time, quasiperiodic evolutions of $\mathcal{C}^{\,}_{zz}(r,t)$ (see Eq.~\ref{eq:czz}) versus time (\emph{b}) and position (\emph{c}) show saturating decay for $\lambda^{\,}_\text{CD}=0.0$ (upper panels), but thermalize from the boundary in for $\lambda^{\,}_\text{CD}=1.0$ (lower panels). 
Results are averaged over $200$ disorder and state realizations for (\emph{a}) and $1000$ realizations for (\emph{b},\emph{c}).
}
\label{fig:CD}
\end{figure*}

However, unlike equilibrium SPTs protected by \emph{microscopic} symmetries, the EDSPT edge modes are obscured both by the frame transformation, $W$, and the quasiperiodic micromotion, $Q(t)$. Since $W$ is a finite depth, static unitary, it merely smears out the edge modes while leaving finite overlap with the 
original edge spins. The time-dependent micromotion, however, encrypts the information encoded in the edge modes in a quasiperiodically rotating frame. The question is then how to recover information stored 
by $\v{\Sigma}^{\,}_{L/R}$ without explicit knowledge of the quasiperiodically rotating frame, $Q(t)$. 

FSPT phases face a similar issue, but offer a simple solution: Since $Q(nT)=Q(0)=\mathbbm{1}$, one can extract the edge state information at integer multiples of the drive period, $T$. In contrast, the \emph{quasi}periodic micromotion never exactly repeats itself: $Q(t)$ does, however, come arbitrarily close to $\mathbbm{1}$ at special ``Fibonacci" times, $t^{\,}_n \sim F^{\,}_n \approx \varphi F^{\,}_{n-1}$, for which $\omega^{\,}_x t^{\,}_n$ and $\omega^{\,}_z t^{\,}_n$ are both exponentially (in $n$) close to integer multiples of $2\pi$; namely, $Q(t^{\,}_n)\approx \mathbbm{1} + \mathcal{O}(\varphi^{-n})$. 

This has two important consequences. First, measuring the edge spin at Fibonacci times allows for the recovery of information with finite fidelity, even at very long times when nontopological bulk modes have fully decohered. Second, since $Q(t)$ quasiperiodically returns (close) to $\ident$ (and $U^{\,}_0(t)$ 
returns precisely to $\mathbbm{1}$), the long-time ``envelope" of the dynamics is effectively controlled by the time-independent Hamiltonian, $\tilde{D}=W^\dagger D W$, which has a pair of \emph{emergent} dynamical symmetries generated by $\tilde{g}^{\,}_{x,z}=W^\dagger g^{\,}_{x,z} W$. We use the term ``emergent'' because (\emph{i}) the precise form of $\tilde{g}^{\,}_{x,z}$ depends on $H(t)$ and (\emph{ii}) arbitrary perturbations to $H(t)$ simply alter the form of $W$ without removing the symmetry. 

Because $\tilde{g}_\mu^2=1$, the corresponding emergent symmetry is $\Z_2\times\Z_2$. The compactness of the emergent symmetry group is essential for the existence of the EDSPT phase, since the group cohomology classification with a pair of integer-time-translation symmetries would be trivial (i.e., $\mathcal{H}^2 \(\Z\times\Z, U(1)\)=\Z_1$).
 The topological protection of the edge can be understood in the usual way: The generators $\tilde{g}^{\,}_{x,z}$ \emph{locally} anticommute acting on the topological edge spin (flipping along either the $x$ or $z$ axis in the $W$ frame), whereas globally, $\[\tilde{g}^{\,}_x,\tilde{g}^{\,}_z\] = 0$. Formally, the emergent symmetry has projective action on the edge modes of $\tilde{D}$. Since these projective representations are discrete, they cannot be continuously changed by perturbations that preserve the structure of Eq.~\ref{eq:hf} (i.e. any sufficiently weak, local, and quasiperiodic perturbation to $H(t)$), which explains the stability of the edge mode dynamics observed in our numerical simulations.

\paragraph{Dynamical anomaly}
An essential characteristic of ordinary $d$-dimensional SPTs is the anomalous, local action of symmetry on the $(d-1)$-dimensional topological edge states, which cannot be implemented in a truly $(d-1)$-dimensional, symmetric system without the accompanying higher-dimensional bulk. Similarly, in Floquet SPTs, every drive period executes an anomalous unitary evolution that cannot be generated by a (symmetric) $(d-1)$-Hamiltonian acting entirely on the edge~\cite{von2016phase,else2016classification,potter2016classification}. These features are essential to the stability of ordinary and Floquet SPTs: Without an anomaly obstruction to realizing the edge symmetry and dynamics, one could apply local perturbations to trivialize the boundary (without breaking any protecting symmetries). 
In-turn, the ability to trivialize the edge would provide a continuous, symmetry-preserving path to deforming the putative SPT to a trivial phase. For example, in the absence of an edge-anomaly obstruction, one could break the system into disconnected pieces, while trivializing the interface between different sections, all-the-while maintaining a (mobility or energy) gap.

This naturally begs the question, what is anomalous about the edge dynamics of the putative EDSPTs model above? or equivalently, is it possible to ``undo" the edge dynamics with a local drive acting purely on the sample boundaries? 
Specifically, we consider applying a ``counter drive" (CD):
\begin{align}
	H^{\,}_{\rm CD} (t) =  -\frac{\lambda^{\,}_{\rm CD}}{2} \sum_{j \in {\rm edge}}\(f^{\,}_x (t)\sigma^x_{j}+f^{\,}_z(t)\sigma^z_{j}\),
\end{align}
to the boundary spins ($A,1$ and $B,L$). For $\lambda^{\,}_\text{CD}= 1$ (perfect $\pi$-pulses), and in the artificial limit where the pulse width vanishes ($\delta$-function pulses), this CD would exactly compensate the putative topological edge dynamics.

However, this $\delta$-pulse limit is incompatible with MBL (or its metastable, prethermal cousin), which requires \emph{smooth} pulses~\cite{ABANIN20161} with limited low- and high-frequency content. For any finite pulse thickness, the CD results in quasiperiodically recurring overlaps between the strong- and non-commuting $x$ and $z$ CD-pulses. Below, we give numerical evidence that these unavoidable pulse-overlaps result in a local transition from quasi-periodic to chaotic, thermalizing dynamics for the counter-driven edge, as the CD strength $\lambda^{\,}_\text{CD}$ is increased beyond a critical value $\lambda^{\ast}_{\rm CD} \sim 0.25$. Furthermore, we find that the chaotic edge thermalizes the entire bulk (loosely analogously to $2d$ static or Floquet MBL systems with a thermal boundary, except in one-lower dimension due to the peculiarity of quasiperiodic systems). This suggests that the edge dynamics of our EDSPTs model exhibits a new form of dynamical anomaly specific to quasiperiodic systems, and that it is not possible to realize a pair of emergent anti-commuting dynamical symmetries by locally driving a $0d$ system.

Fig.~\ref{fig:CD}a shows the evolution of the edge correlations as a function of counter drive strength. As a baseline, we note that, after a short transient, the bulk correlation functions exhibit random-disorder dependent oscillations, whose average value decays to zero with the number of disorder configurations $N_\text{dis}$ as $\sim 1/\sqrt{N_\text{dis}}$. Without the CD, the disorder-averaged edge correlation plateaus at a nonzero, $N_\text{dis}$-independent value, indicating the presence of a topological edge mode (with nonzero overlap with the end spin) that is dynamically decoupled from the local disorder. Turning on a weak CD ($\lambda^{\,}_{\rm CD} \lesssim 25\%$) gradually reduces the value at the plateau value without destroying its presence. 

For stronger drives, up to $\lambda_\text{CD}^*\equiv 0.25\lesssim \lambda^{\,}_\text{CD} \leq 1$ the topological protection of the edge mode is destroyed, and the CD gives vanishing disorder-averaged edge correlations. These behaviors are separated by a characteristic CD strength $\lambda_\text{CD}^* \approx 0.25$. However, the destruction of correlations is not confined to the system boundary. To explore the bulk behavior, we examine correlation functions: 
\begin{align}
	\mathcal{C}^{\,}_{zz}(r,t) =  \overline{\left\vert \<n|\Sigma^z_r(t)\Sigma^z_r(0)|n\>\right\vert~},
\end{align}
 of the LIOMs of $H^{\,}_\text{CS}$, averaged over a number of disorder realizations, starting from a different random $\sigma^z$-product state for each realizations (note that absolute values are taken to prevent cancellation of oscillatory terms with disorder averaging), where:
\begin{align}
	\Sigma^z_r = 
	\begin{cases}
		\sigma^z_{A,1}, & r=1\\
		\sigma^z_{B,r/2}\sigma^z_{A,r/2+1}, & r~\text{even}, 1<r<2L \\
		\sigma^z_{B,L}, & r=2L
	\end{cases}
	\label{eq:czz}
\end{align}
which have non-negligible overlap with the emergent LIOMs of the quasiperiodic system in the absence of the CD. Here, $r$ indexes position along the spin-chain (without regard to the $A/B$ sublattice structure), and $\Sigma^z_{r=1,L}$ correspond to topological edge-spin operators for $H^{\,}_\text{CS}$, whereas the remainder correspond to bulk LIOMs. We observe that the for $\lambda^{\,}_\text{CD}> \lambda_\text{CD}^*$, both the bulk and edge correlators $\mathcal{C}^{\,}_{zz}(r,t)$ eventually decay to zero (instead of saturating as for $\lambda^{\,}_\text{CD}< \lambda_\text{CD}^*$), suggesting that both the bulk and boundary are thermalizing. Moreover, by examining the spatial dependence of $\mathcal{C}^{\,}_{zz}(r,t)$ for different times (Fig.~\ref{fig:CD}(b,c) ), one clearly observes that bulk spins thermalize later than edge spins, with the thermalization time increasing with distance into the bulk. This suggests that $\lambda_\text{CD}^*$ marks a boundary phase transition between quasiperiodic and chaotic edge dynamics, with the chaotic edge-spin serving as a continuous-spectrum noise-source that thermalizes the bulk.

\paragraph{Boundary thermalization in Floquet approximants}
Unfortunately, due to the absence of any conserved energy or well-defined eigenstates in quasiperiodic systems, ordinary metrics of thermalization can not be analyzed. To better assess this boundary CD thermalization scenario, we instead introduce a sequence of Floquet proxies for the quasiperiodic drive, wherein we replace $T^{\,}_{z} = \varphi$ with a rational approximant of $\varphi \approx F^{\,}_{n+1} / F^{\,}_n$, with $T^{\,}_{x}=1$, resulting in overall periodicity with period $T = F^{\,}_{n+1}$. During each period, the $x$-pulse is applied $F^{\,}_{n+1}$ times and the $z$-pulse $F^{\,}_n$ times. In the limit $n \to \infty$, $T^{\,}_{z}  \to \varphi$, and the system becomes truly quasiperiodic ($T \to \infty$). By examining a sequence of finite-$n$ approximants, we numerically probe the level statistics of the Floquet evolution operator and half-chain entanglement entropy of its eigenstates to diagnose thermalization versus MBL, and examine both infinite time correlations and stroboscopic evolution of correlation functions.

Before discussing the numerical results, it is worth pausing to consider the relation between the Floquet-approximants and true quasiperiodic drive. The $n$th Floquet approximant drive approximately agrees with the quasiperiodic evolution up to time $t\sim F_n$. However, the eigenstates of the Floquet approximant reflect \emph{infinite}-time behavior for times well-beyond $t\sim F_n$ where the drives no longer (even approximately) agree. Despite this, we claim that the \emph{localization} properties of the Floquet eigenstates predict those of the quasiperiodic drive. Specifically, if the quasiperiodic system is MBL, and has an extensive set of LIOMs, then within time $t\sim F_n$, it is possible to approximately construct these LIOMs by time-averaging local operators~\cite{PhysRevB.91.085425}, with error $\sim 1/\text{poly}(t)$. Hence, if each of the Floquet approximants is MBL, then the LIOMs will converge as $n\rightarrow \infty$, and will coincide with the LIOMs of the fully-quasiperiodic drive. In contrast, if the approximants thermalize, then this implies that the quasiperiodic drive also thermalizes. We caution though, that while the localization properties of the approximants extend to the quasiperiodic drive, the \emph{stroboscopic} dynamics of the Floquet approximants beyond the first period are not directly related to the quasiperiodic time-evolution.

\begin{figure}[t]
\centering
\begin{overpic}[width=1.0\columnwidth]{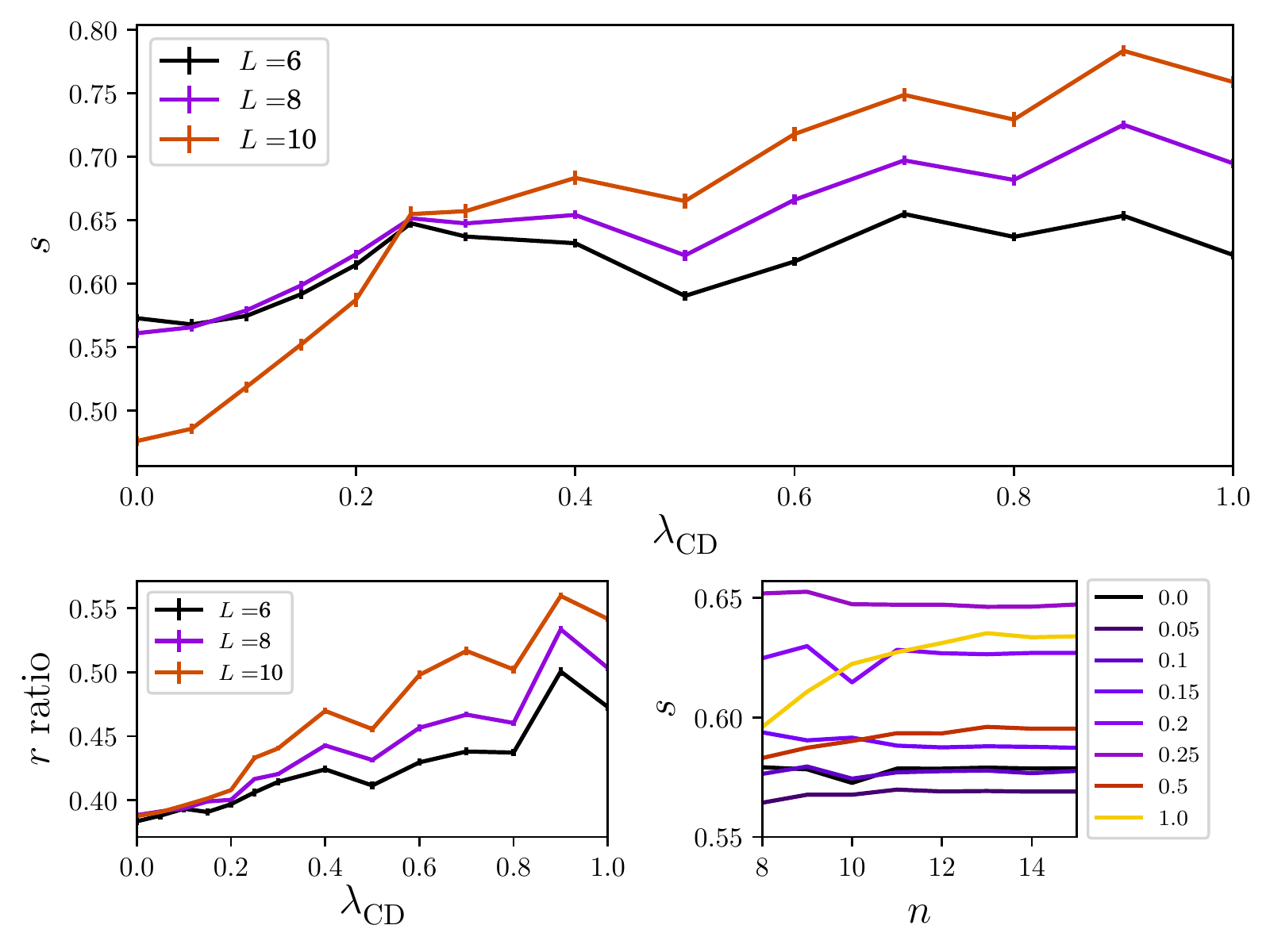} 
 \put (0,35) {(a)}
 \put (0,2) {(b)}
  \put (52,2) {(c)}
\end{overpic}
\caption{{\bf Boundary thermalization in Floquet Approximants---}
Spectral signatures of localization and thermalization for the $n=10^\text{th}$ Floquet approximant to the quasiperiodic drive. (\emph{a}) Finite size crossing in normalized half-system entanglement at a critical edge counter drive (CD) strength, $\lambda_\text{CD}^*\approx 0.25$, which divides localized and thermal regimes. (\emph{b}) The corresponding behavior in level-statistics. (\emph{c}) The $n$ dependence of these quantities saturates for large $n$, and is well-saturated by $n=10$ near $\lambda_\text{CD}^*\approx 0.25$. 
Each result reflects an average over the full spectrum for $240$ disorder realizations.
}
\label{fig:floq_approx}
\end{figure}

Fig.~\ref{fig:floq_approx} shows the $r$-ratio~\cite{PhysRevB.75.155111} for the Floquet quasi-energy spectrum and the (normalized) half-system eigenstate entanglement entropy, $s=S/L$ (taken ${\rm log}^{\,}_2$ and with $L$ being half the total number of sites) for Floquet approximants. In the absence of a CD, we observe MBL-like behavior (r-ratio close to Poisson, small entanglement) for all $n$ and $L$, providing evidence that our model is indeed in the MBL regime. Observing that, the $n$-dependence of these quantities quickly saturates, we henceforth concentrate on a single approximant, $n=10$, and turn to the effect of the CD. Though the system sizes are admittedly limited (due to the long time-integration needed to construct each Floquet approximant), we observe evidence that the half-system entanglement, $s$ exhibits a finite-size crossing from MBL (area-law) to thermal (volume-law) behavior  by $\lambda_\text{CD}^* \approx 0.25$, signaling a potential thermalizing phase-transition at this CD-strength, consistent with the correlation function observations for the full quasiperiodic drive. Similarly, the r-ratios show Poisson-like behavior below $\lambda_{\rm CD}^*$, and then change towards chaotic for stronger drives $\lambda^{\,}_\text{CD}>\lambda_\text{CD}^*$, but do not show a sign of a finite-size crossing. We attribute this unconventional scaling behavior to the unusual boundary-thermalization nature of this transition, in which the thermalization is induced purely from a transition in the edge-spin dynamics from quasiperiodic to chaotic, which then acts as a continuous-spectrum noise source that melts the bulk MBL. Consequently, we do not expect to see conventional scaling of $r$ with $L$, since adding additional bulk MBL degrees of freedom does not effect the boundary criticality (note that the scaling behavior in $s$ is explained by the trivial $\sim 1/L$ normalization, which trivially causes a finite-size scaling of $s \sim 1/L$ in the area-law regime).

We note that, while our numerical observations are consistent with the scenario in which a $0d$, local criticality of the edge spin thermalizes the bulk, the achievable system-sizes are somewhat limited, and we are unable to rule out that this trend is ultimately a finite-size artifact. For example, it could be that the CD makes a moderate- but finite-size (e.g. 5-6 spins) thermal ``puddle" near the end of the chain, which does not ultimately spread and drive a bulk phase transition. To this end, analytic insight into the ultimate fate and nature of the boundary-thermalization transition are highly desirable targets for future inquiry.

\paragraph{Discussion}
To summarize, we have constructed a two-tone quasi-periodic driving protocol that produces a pair of long-lived dynamically protected edge-modes, that are (at least) exponentially insensitive to generic perturbations without any symmetry constraints. Whether this results in a long-lived but ultimately metastable pre-heating phenomena, or a genuine infinitely long-lived phase is a challenging but interesting question for future theoretical work. However, we note that, in practice, this issue is likely to be a largely academic question, given that preheating times can vastly exceed finite experimental lifetimes over a wide range of parameters. Moreover, our numerical evidence is consistent with the scenario that this is a genuine MBL phase, with non-trivial topological edge dynamics that cannot be removed without a bulk phase transition. 

Apart from its dynamically protected edge-state phenomenology, perhaps the most intriguing implication of this example is that it is missing from previously conjectured topological classification schemes. Na\"ively, one could also attempt to apply our construction to produce Floquet EDSPTs. Specifically, starting from an exactly solvable model realizing an SPT with symmetry group: $G\times \Z^{\,}_{n_1}\times\dots \Z^{\,}_{n_N}$ (this plays the role of $H^{\,}_\text{CS}$ above), one could attempt to replace the microscopic symmetry $G$ with a corresponding emergent dynamical symmetry by applying an N-tone quasi-periodic pulse train a la $H^{\,}_0(t)$ but replacing $g^{\,}_{x,z}$ with the generators of the $\Z_{n_1}\times\dots \Z_{n_N}$ symmetry factors. The minimal example would be a $2d$ EDSPTs version of the Levin-Gu SPT phase whose single Ising ($\Z_2$) symmetry was traded for an emergent dynamical symmetry enforced by a periodic drive. However, in App.~\ref{app:FloqCD}, we show that this model can actually be trivialized by applying an appropriate counter-drive to undo the edge motion, \emph{without} causing a thermalization transition (unlike the quasiperiodic example above). Based on these examples, we conjecture that EDSPTs are special to quasiperiodic systems. 

\vspace{4pt}\noindent{\it Acknowledgements -- } We thank P. Dumitrescu for helpful discussions and closely related prior collaboration. This work was supported by DMR-1653007 (ACP),  the US Department of Energy, Office of Science, Basic Energy Sciences, under Early Career Award No. DE-SC0019168 (RV), and the Alfred P. Sloan Foundation through
a Sloan Research Fellowship (RV). Numerical diagonalization simulations were performed on the Texas Advanced Computing Center (TACC).

\bibliography{qpdc_bib}
\newpage
\onecolumngrid
\appendix

\section{Quasiperiodic driving and high-frequency expansion} \label{app:ElseDetails}
\subsection{Cut and project method} \label{app:cutandproject}
It is often useful to view a quasiperiodic function as a projection of a ``slice" through a higher-dimensional \emph{periodic} function: A two-tone quasiperiodic drive can be Fourier-expanded as $H(t,\vec{\phi}) = \sum_{\vec{n}\in\Z^2}e^{-i\vecn  \cdot \(\vecomega t+\vec{\phi}\)} \, H^{\,}_{\vecn}$ where $\omega^{\,}_1 / \omega^{\,}_2 \notin \mathbb{Q}$; consequently, we can view $H (t)$ by evaluating a periodic function of a two-dimensional vector, $\vec{\theta}$, projected onto the trajectory $\vec{\theta}(t, \vec{\phi})=(\vec{\omega}t+\vec{\phi})$, i.e.,
\be H(t,\vec{\phi})\equiv H\[\vec{\theta}(t, \vec{\phi})\]	~~,~~ \ee
where, in a slight abuse of notation, we use $H$ to refer to both of the equivalent $t$ and $\vec{\theta}$ parameterizations. The drive,
$H(\vec{\theta})$, is periodic under two independent time translation ``symmetries": $\vec{\theta}\rightarrow \vec{\theta}+2\pi\uvec^{\,}_\mu$, where $\uvec^{\,}_\mu$ is a unit vector in the `$xz$'-plane (for $N$-tone drives, $\vec{\omega},\vec{n},\vec{\theta}$ generalize to $N$-component vectors). 

In general, the resulting time evolution, $U(t)$, is \emph{not} quasiperiodic in $t$. However, in many instances, a multi-mode extension of Floquet's theorem applies~\cite{jorba1992reducibility} and allows $U(t)$ can be reduced to a quasiperiodic modulation accompanied by a static Hamiltonian evolution, and several techniques~\cite{xue2018effective,else2020long} have been introduced to approximately construct the effective Hamiltonian in the weak driving or high frequency limit.

Such perturbative methods do not directly apply to our model, due to the requirement of strong pulses with weight near $\pi$.
A recent work~\cite{else2020long} shows that this obstacle can be circumvented for drives that are sufficiently close to a solvable limit (e.g., $H^{\,}_0(t)$ as defined in Eq.~\ref{eq:H0def}), by first transforming into the interaction picture of $H^{\,}_0(t)$ to exactly account for the strong part of the dynamics. Then an appropriate high-frequency expansion can be performed in this rotating frame. For these cases, $U(t)$ can be broken down into the form shown in Eq.~\ref{eq:hf}, consisting of a time-independent MBL evolution, $D$ (which simply rotates localized bulk degrees of freedom by an overall phase), and a quasiperiodic micromotion generated by $Q$ and $U_0$. An important caveat is that, while $Q$ has the same quasiperiodicity as $H$, $U_0(t)$ has a doubled periodicity: 
\begin{align}
	Q(\vec{\theta}+2\pi\uvec_\mu) &= Q(\vec{\theta})
	\label{eq:QTTS} \\
	U_0(\vec{\theta}+2\pi\uvec_\mu) &= U_0(\vec{\theta}) \, g^{\,}_\mu,
	\label{eq:ttts}
\end{align}
where a single ``time translation'' about the $\mu=x,z$ axis has the effect of transforming the system by $g^{\,}_\mu$, which is a symmetry of the effective quasi-Floquet Hamiltonian, $D$. Eq. \ref{eq:ttts} is referred to as ``twisted-time translation" symmetry in Ref.~\cite{else2020long}.

In this sense, the emergent dynamical symmetry can be thought of as arising from (the projection of) a multi-time translation symmetry, with an independent time ``direction'' ($\uvec^{\,}_{x,z}$) for each incommensurate tone of the drive ($\omega^{\,}_{x,z}$).

\subsection{High-frequency expansion} \label{app:HFE}
Here we briefly review the interaction picture high-frequency expansion approach developed in~Ref.~\citenum{else2020long} to compute $W,Q,D$ in Eq.~\ref{eq:hf} approximately (i.e., to some specified order). We illustrate this approach for the model described in the main text. The first step is to split the full quasiperiodic Hamiltonian $H(t)=H^{\,}_0(t)+H'(t)$, into the ideal (unperturbed) drive, $H^{\,}_0(t)$, and the remaining terms $H^{\prime}_S = H^{\,}_\text{CS}+V(t)$, and transform the Schr\"odinger picture  Hamiltonian, $H'_S$, into the interaction picture of $H^{\,}_0$:
\begin{align}
	H^{\prime}_{\rm int} (t) = U^{\dagger}_0 (t) \, H^{\prime}_{\rm S} (t) \, U^{\,}_0 (t).
\end{align}
This interaction frame Hamiltonian inherits the enlarged quasiperiodicity of $U^{\,}_0$  (see Eq.~\ref{eq:ttts})---i.e., in the time-torus parameterization, we have $H^{\prime}_{\rm int}(\vec{\theta}+4\pi \uvec^{\,}_\mu)=H^{\prime}_{\rm int}$.

We can write the full evolution operator, $U (t)$ as $U (t) = U^{\,}_0 (t) \, U^{\,}_{\rm int} (t)$, where $U^{\,}_{\rm int}$ satisfies $i \, \partial^{\,}_t \, U^{\,}_{\rm int} (t)  = H^{\prime}_{\rm int} (t) \, U^{\,}_{\rm int} (t)$, and where all terms can be regarded as functions of $\vecth (t)$. The goal will be to identify a quasiperiodic frame transformation, $P\(\vec{\theta}(t)\) = e^{-i\Gamma(\vec{\theta})}$, that reduces $U^{\,}_\text{int}$ to an effective, time-independent Hamiltonian evolution, $e^{-iDt}$, i.e.,
\be \label{eq:newUint} 
	U^{\,}_{\rm int} \[ \vecth (t) \] \equiv P \[ \vecth (t) \]  e^{-iDt}  \underset{\equiv W}{\underbrace{P^{\dagger} \[ \vecth (0) \]}},
\ee
where we define the $t=0$ frame rotation operator, $W$, for convenience. 

The operator $P$ is not $2\pi$-periodic in $\theta^{\,}_{x,z}$; rather, $2\pi$ shifts in the components of $\vecth$ conjugate $P$ by the corresponding emergent symmetry: $P(\vec{\theta}+2\pi\uvec^{\,}_\mu)= g^{\,}_\mu P(\vec{\theta}) \, g^{\,}_\mu$~\cite{else2020long} (i.e., $P$ is covariant under the time-translation symmetries). 

We find it convenient to depart from the conventions of Ref.~\cite{else2020long}, instead regrouping terms to define a quasiperiodic micromotion operator, as in Eq.~\ref{eq:hf},
\be 
	Q \equiv W \,  U^{\,}_0 \, P \, U_0^\dagger 
	~.~~ \label{eq:Q} 
\ee 
Note that the noninvariance of $P$ under $2\pi$ shifts of $\vec{\theta}$ is precisely compensated by the inverse behavior in $U^{\,}_0$, so that the micromotion, $Q$, is $2\pi$-periodic in both components of $\vecth$ (i.e., $Q$ has the original quasiperiodicity of $H(t)$, rather than the enlarged, ``twisted" quasiperiodicity of $U^{\,}_0$).

\subsection{Effective quasi-Floquet Hamiltonian} \label{app:Heff}
For drives close to $H^{\,}_0$---i.e., those for which $T^{\,}_x H^{\prime}_{\rm int}$ is small---may be treated perturbatively in a high-frequency (or equivalently weak-coupling) approximation. Denoting the size of the local Hamiltonian terms scaled by $T_x$ as $v$, Ref.~\citenum{else2020long} derives expressions for $D$ and $P$ (reproduced in Eq.~\ref{eq:newUint}) order-by-order in $v$ in terms of nested commutators of Fourier components,
\be \label{eq:FourierComp} 
	H^{\prime}_{\vecn} =\iint_{0}^{4\pi} \frac{d^2\vec{\theta}}{(4\pi)^2}\, e^{-i \, \vecn \cdot \vecth /2} \, H^{\prime}_{\rm int} \( \vecth \) 
\ee 
of the interaction-frame perturbation terms. The enlarged range of integration accounts for the doubled time translation symmetry (i.e., $4 \pi \vecsym{e}^{\,}_{\mu}$ and not $2 \pi \vecsym{e}^{\,}_{\mu}$). We use this form for notational convenience: One could alternatively implement a change of variables on $\vecth$ or allow half integer $\vecn$ to recover a more typical expression. 

These Fourier components define the quantities
\begin{align}
	D = \sum_{q=1}^\infty D^{(q)}, ~~~~
	P = e^{-i\Gamma} = \exp\[-i\sum_{q=1}^\infty \Gamma^{(q)}\],
\end{align} 
where the $q$th term is of size ${\cal O}(v^{q+1})$, and solving order by order in $v$ in Eq.~\ref{eq:newUint} recovers the expressions for the components at each order~\cite{else2020long}, as we show below.

\paragraph{Effective Hamiltonian}
The contributions to the effective time-independent Hamiltonian, $D^{(q)}$, are obtained by considering
\be \label{eq:Dgen} 
	D = P^{\dagger} (t) \, H^{\prime}_{\rm int} (t) \, P (t) - i \, P^{\dagger} (t) \, \partial^{\,}_t \, P (t),\ee
order by order in $v$, and demanding that $D$ be independent of $\vec{\theta}(t)$.

The leading two terms closely resemble those of the Magnus expansion for Floquet systems, 
\begin{align}
	D^{(1)} &= H^{\prime}_{\vec{n}=0}
	\label{eq:D1} \\
	D^{(2)} &= \sum_{\vecn \in \Z^2 \neq 0} \frac{1}{2 \vecomega \cdot \vecn} \[H^{\prime}_{\vecn},H^{\prime}_{-\vecn}\]~,~~\label{eq:D2}
\end{align}
and the $q$th correction, $D^{(q)}$, comprises $q$ nested commutators of $H^{\prime}_{\vecn^{\,}_j}$, subject to the condition $\sum_{j=1}^{q} \vecn^{\,}_j = 0$. The fact that the Fourier indices sum to $0$ is necessary and sufficient for $D$ to be static, as we show in App. \ref{app:emergesymprop}.

The leading term, $D^{(1)}$, is simply the average value of $H'_\text{int}$. For the model described in the main text, $D^{(1)} \sim H^{\,}_\text{CS} + \sum_{i=1}^L \sum_{\nu=x,z} J_i^\nu \sigma_{A,i}^\nu\sigma_{B,i}^\nu$, which, for $J<K$, is in an AKLT/cluster-state phase. Notice that the single-spin field terms $\sim \vec{h}\cdot\vec{\sigma}$ drop out of $D^{(1)}$, as they are twirled over the emergent symmetry group upon computing the average over $\vec{\theta}$. General expressions for higher-order terms quickly become cumbersome, and are not particularly illuminating, other than to note that they all necessarily commute with  $g^{\,}_{x,z}$ (as shown in App. \ref{app:emergesymprop}), and come with small coefficients that are appropriately suppressed by powers of $K,h,\dots$, and die off rapidly with $\abs{\vecn} > 1$.

\paragraph{Micromotion}
The leading order contributions to the generator of micromotion are
\begin{align}
\Gamma^{(1)}(\vec{\theta}) &=  \sum\limits_{\vec{n} \in \Z^2 \neq 0} \frac{e^{i \vecn \cdot \vecth /2} }{i \vecomega \cdot \vecn} H^{\prime}_{\vecn } \label{eq:Gamma1} \\ \Gamma^{(2)}(\vec{\theta})  &= \sum\limits_{\substack{ \vecn \in \Z^2 \neq 0 \\ \vecm \neq \vecn }} \frac{e^{i \vecn \cdot \vecth/2} }{i \vecomega \cdot \vecn} \frac{1 + \delta^{\,}_{\vecm, 0} }{2 \, \vecomega \cdot \( \vecn - \vecm \)} \, \[ H^{\prime}_{\vecn - \vecm} , H^{\prime}_{\vecm} \]  \label{eq:Gamma2}
\end{align}
Using Eq.~\ref{eq:Q}, this implies:
\begin{align}
 Q\(\vec{\theta}\) &= e^{i\Lambda} = e^{i\sum_{n}\Lambda^{(n)}}
 \nonumber\\
 \Lambda^{(1)} &= U^{\,}_0\Gamma^{(1)}U_0^\dagger - \Gamma^{(1)}(0)
  \nonumber\\
 \Lambda^{(2)} &= U^{\,}_0\Gamma^{(2)}U_0^\dagger - \Gamma^{(2)} (0) -\frac{i}{2}\[U^{\,}_0\Gamma^{(1)}U_0^\dagger,\Gamma^{(1)}(0)\]
 ,
\end{align}
where $\Gamma^{(q)}$ is evaluated at $\vecth (t)$ unless otherwise stated.

\paragraph{Terms for the AKLT model} 
Most of the terms in $H^{\prime}_S = H^{\,}_\text{CS}+V(t)$ (where the full Hamiltonian is given by $H(t)=H^{\,}_0(t)+H'(t)$) are modified by shifting to the interaction picture of $H^{\,}_0(t)$. However, two of the terms in $V(t)$ commute with all terms in $H^{\,}_0$, and their Fourier components can be found analytically.

For the random $AB$ terms, $\sum_j \, \sum_{\nu=x,z} J^{\nu}_j \sigma^{\nu}_{A,j} \sigma^{\nu}_{B,j}$ (where we take the couplings to be time independent for simplicity), we have from Eq.~\ref{eq:FourierComp},
\be H^{\prime}_{\vecn} = \delta^{\,}_{\vecn \, , \, 0} \sum\limits_{j=1}^L \, \sum_{\nu=x,z} J^{\nu}_j \sigma^{\nu}_{A,j} \sigma^{\nu}_{B,j}~,~~\ee
i.e., there is no change to this term and only the $\vecn = 0$ term is nonzero.

The corrections to the $H^{\,}_0(t)$ pulses (i.e., deviation from a $\pi$ pulse) can also be computed exactly. Again, there is no change going to the interaction picture, and the correction to the pulse $\nu=x,z$ has the form
\be H^{\prime}_{\vecn} =\frac{ \lambda }{4 \pi}  \sum\limits_{j=1}^L \, \sum_{\nu=x,z} \omega^{\,}_{\nu}\, e^{-w^2 n^{2}_{\nu}/8}  \sigma^{\nu}_{A,j} \sigma^{\nu}_{B,j}~.~~\ee
if $n^{\,}_{\nu}$ is even and $n^{\,}_{\overline{\nu}} = 0$, and is zero otherwise; these terms fall off as $e^{-\kappa n^{2}}$.

The other terms in $H^{\prime}_S$ (i.e., $H^{\,}_\text{CS}$ and the random fields) are modified upon going to the interaction picture, where they show nontrivial time dependence. Because of this, their Fourier coefficients can only be evaluated numerically, though they still appear to fall off at least exponentially in $n$ (Fourier components for $n^{\,}_{\nu} \gtrsim 20$ are zero to numerical precision, and decay faster than $2^{-n}$ for the parameters used for numerical simulation).

The random $x$ fields $H^{\prime}_{\rm S} = h^{x}_{\alpha,j} \sigma^{x}_{\alpha,j}$ (where $\alpha = A,B$ labels the sublattice), upon going to the interaction picture of $H^{\,}_0(t)$ become
\begin{align} \label{eq:xfieldintpic} H^{\prime}_{\rm int} &= h^{x}_{\alpha,j} \left( \cos \left[ F^{\,}_z (t) \right] \, \sigma^{x}_{\alpha,j} - \sin \left[ F^{\,}_z (t) \right] \, \sigma^{y}_{\alpha,j} \sigma^{z \vphantom{y}}_{\overline{\alpha},j}\right) ~,~~~~\end{align}
where $\overline{A} = B$ and vice versa, and
\be F^{\,}_{\nu} (t) = \int\limits_0^{t} ds \, f^{\,}_{\nu} (s)~,~~\ee
with $f^{\,}_{\nu} (s)$ the Gaussian pulse defined in Eq.~\ref{eq:H0def}.

Similarly, for the random $z$ fields,  $H^{\prime}_{\rm S} = h^{z}_{\alpha,j} \sigma^{z}_{\alpha,j}$, going to the interaction picture gives
\begin{align} \label{eq:zfieldintpic} H^{\prime}_{\rm int} &= h^{z}_{\alpha,j} \left( \cos \left[ F^{\,}_x (t) \right] \, \sigma^{z}_{\alpha,j} + \sin \left[ F^{\,}_x (t) \right] \, \sigma^{y}_{\alpha,j} \sigma^{x \vphantom{y}}_{\overline{\alpha},j}\right) ~,~~~~\end{align}
and random $y$ fields, $H^{\prime}_{\rm S} = h^{y}_{\alpha,j} \sigma^{y}_{\alpha,j}$, which fail to commute with both pulses in $H^{\,}_0$ are more complicated:
\begin{align}  H^{\prime}_{\rm int} &= h^{y}_{\alpha,j} \left( c^{x}_t c^{z}_t \, \sigma^{y}_{\alpha,j} + s^{x}_t s^{z}_t \, \sigma^{y}_{\overline{\alpha},j}  \right. \notag \\
&~~~+ \left. c^{x}_t s^z_t \sigma^{x}_{\alpha,j}  \sigma^{z}_{\overline{\alpha},j} - s^{x}_t c^z_t \sigma^{z}_{\alpha,j}  \sigma^{x}_{\overline{\alpha},j}     \right) \label{eq:yfieldintpic}~,~~~~\end{align}
where $c^{\nu}_t$ is a shorthand for $\cos \left[ F^{\,}_{\nu} (t) \right]$ (and $s^{\nu}_t$ for $\sin \left[ F^{\,}_{\nu} (t) \right]$).

The $x$ field terms are zero unless $n^{\,}_z$ is odd and $n^{\,}_x = 0$; the $z$ field terms are zero unless $n^{\,}_x$ is odd and $n^{\,}_z = 0$; the $y$ field terms are zero unless $n^{\,}_{x,z}$ are \emph{both} odd. We demonstrate this property analytically in App.~\ref{app:emergesymprop}. 

The interaction picture form of the stabilizer terms, $ H^{\,}_\text{CS}$, can be recovered from Eqs.~\ref{eq:xfieldintpic} and \ref{eq:zfieldintpic}. Each term $ K^\nu_{j} \sigma^\nu_{B,j}\sigma^\nu_{A,j+1}$ contains $\sigma^{\nu}$ terms in two neighboring unit cells; going to the interaction picture results in four terms: for the Schr\"odinger term $\sigma^x_{B,j}\sigma^x_{A,j+1}$, the dominant term in the interaction picture is of the same form, $\sigma^x_{B,j}\sigma^x_{A,j+1}$; other (smaller) corrections include $\sigma^x_{B,j}\sigma^y_{A,j+1}\sigma^z_{B,j+1}$, $\sigma^z_{A,j}\sigma^y_{B,j}\sigma^x_{A,j+1}$, and $\sigma^z_{A,j}\sigma^y_{B,j}\sigma^y_{A,j+1}\sigma^z_{B,j+1}$. Similar terms emerge for the $K^z_{j}$ terms. Because the stabilizer terms commute with $g^{\,}_{x,z}$, they are nonzero only for even Fourier indices. 

When the pulse width ($w$) is much smaller than the period (e.g., $w = 0.05 T$ as used for numerical simulation), the Fourier transforms of the cosine terms above are roughly $0.9$ to $0.95$ (for the smallest Fourier coefficients, $n=0,$), and the sine terms are roughly $0.05$ to $0.1$. Subsequent Fourier coefficients ($n \gg 0$) will be exponentially smaller. Regarding Eqs.~\ref{eq:xfieldintpic} through \ref{eq:yfieldintpic}, it is apparent that narrow pulses minimize the ``new'' terms (i.e., those different from the Schr\"odinger picture form of the operators), and most of the physics can be understood from the Schr\"odinger form of the operator and the suppression in corrections to $D$. This holds for both the field and stabilizer (cluster) terms.

\paragraph{AKLT Effective Hamiltonian} We can now examine the contribution of the terms in $H^{\prime}_{\rm int} $ to $D$, starting with the lowest order terms, $D^{(1)}$. 

This term consists of the $\vecn = 0$ components of $H^{\prime}_{\vecn}$. First, we have the intracell $AB$ terms, $\sum_j \, \sum_{\nu=x,z} J^{\nu}_j \sigma^{\nu}_{A,j} \sigma^{\nu}_{B,j}$, exactly as they appear in the Schr\"odinger picture. Additionally, we have a contribution from the pulse correction, $ \lambda /2  \sum_{j=1}^L \, \sum_{\nu=x,z} T^{-1}_{\nu} \, \sigma^{\nu}_{A,j} \sigma^{\nu}_{B,j}$, where $\lambda$ captures the deviation from a $\pi$ pulse. The field terms are zero for $n^{\,}_x = n^{\,}_z = 0$. The final contribution to $D^{(1)}$ comes from the stabilizer terms, and for $w=0.05T$ (the value used for numerical simulations), the primary contribution is roughly $0.9 \times \sum_k \sum_{\nu} K^\nu_{j} \sigma^\nu_{B,j}\sigma^\nu_{A,j+1}$ (i.e., $90 \%$ of the bare Schr\"odinger term), plus corrections spanning both $j$ and $j+1$ with prefactors of roughly $0.1 K^\nu_{j}$. 

For $K \gg J$, $w \ll 1$, and small deviation, $\lambda$, from a $\pi$ pulse, this $D^{(1)}$ will correspond to a Hamiltonian in the AKLT phase. Increasing the strength of the $J$ couplings or the deviation, $\lambda$ from a $\pi$ pulse, or decreasing the strength of the cluster terms, $K$, can result in $D^{(1)}$ realizing the trivial phase.
 
The next order correction, $D^{(2)}$, consists of sums over commutators of $H^{\prime}_{\vecn}$ and $H^{\prime}_{-\vecn}$, for $\vecn \neq 0$. The intracell terms (with coefficients $J{\nu}_j$) do not contribute, as they only have $\vecn=0$ coefficients. However, the field terms, $h^{\mu}_{\alpha, j} \sigma^{\mu}_{\alpha, j}$, \emph{do} contribute to  $D^{(2)}$.

However, the contribution of the field terms is limited. Firstly, the symmetry restrictions mean that the only terms entering the summand in Eq.~\ref{eq:D2} are of the form $ h^{\mu}_{\alpha, j} \, h^{\mu}_{\alpha^{\prime}, j}$ (i.e., same type of field and acting on the same cell). The $x$ field terms, e.g., generate terms of the form $\sigma^{z}_{A, j}\sigma^{z}_{B, j}$ and $\sigma^{y}_{A, j}\sigma^{y}_{B, j}$, which have an effect similar to the $J^{\nu}_{j}$ terms. However, summing over Fourier coefficients results in an overall suppression of $O(10^{-2})$, in addition to the small perturbative factor of $O(h^2)$. Hence, for small fields, $h$, these terms are not particularly harmful on their own.

The remaining terms have only even Fourier components, and are somewhat restricted in that most terms have one of $n^{\,}_{x,z}$ zero (with the other even). The pulse corrections do not produce new terms on their own. The stabilizer terms, $K^{\nu}_{j}$ produce new intercell terms, which may act like the original $K$ terms or as more complicated hopping or interaction terms (in terms of the cluster LIOMs). Additionally, the stabilizers and pulse correction will produce additional such terms. 

However, due to the number of terms and inability to compute their coefficients analytically, we resorted to constructing $D$ to second order numerically. For the parameters used for numerical simulation, we find exact commutation of $D$ with $g^{\,}_{x,z}$ (to numerical precision), Poisson statistics, and edge modes. This is further supported by time evolution and numerical diagonalization of Floquet rational approximants of the quasiperiodic drive.

\paragraph{Convergence} \label{app:HFEconverge}
As in the Floquet-Magnus expansion, the $q$th order terms in $D$ and $\Gamma$ are each suppressed by $\sim v^q$, but grow in number combinatorially as $\sim q!$. Thus, the expansion is asymptotic---rather than truly convergent---and should be truncated to some optimal order, with weight of truncated terms $\sim te^{-1/v^\gamma}$ (with $\gamma \lesssim 2/3$, see below), indicating that the approximations become inaccurate for $t\gtrsim e^{1/v^\gamma}$; beyond this time, the expansion is not necessarily predictive~\cite{else2020long}. In strongly disordered Floquet systems there is numerical evidence that stable MBL can persist beyond the time scale set by the asymptotic high-frequency expansion, at least in $1d$ (and possibly also higher $d$, either ignoring rare thermal region effects, or in the case of spatially quasiperiodic ``disorder"). However, analytical evidence of such stability remains elusive.

In addition to these concerns, unlike the Floquet expansions, in the quasiperiodic setting one also must consider small denominators, $\vecomega \cdot \vecn \sim 0$, which occur for rational approximates of the ratio of the base periods. For our model with $\omega^{\,}_{x}/\omega^{\,}_{z} = \varphi$, this occurs for $\vec{n}$ given by successive Fibonacci numbers, i.e., $\vecn^{\,}_k = (F^{\,}_k, -F^{\,}_{k-1})$, such that $\vecomega \cdot \vecn^{\,}_k = F^{\,}_k -\varphi \, F^{\,}_{k-1} \propto \varphi^{-k}$. Generally, accurate convergence of the expansion requires that the numerator of these terms decays sufficiently rapidly with $\abs{\vecn}$. 

For the Gaussian-pulse model presented above, all Fourier amplitudes decay $\sim e^{-n_k^2}\sim e^{-\varphi^{2k}}$, which tend to zero much more quickly with $k$ than $\vec{n}_k\cdot\vec{\omega}\sim \varphi^{-k}$. Additionally, terms that commute with the $x$ [$z$] pulse necessarily have $n^{\,}_x = 0$ [$n^{\,}_z = 0$]; thus, only perturbations that fail to commute with \emph{both} pulses pose a risk in the sense of small denominators. Following the logic of Ref.~\cite{else2020long} for Gaussian pulses, and assuming that the system can re-arrange itself to absorb the energy from the drive (which is not the case if the system is MBL), we find a heating time scale $t\sim e^{(K/v)^{\gamma}}$ with $\gamma \lesssim 2/3$. Note that the actual heating time scale is potentially much larger in the presence of strong disorder, and possibly infinite if the system is truly many-body localized. 

The most natural perturbation of this type corresponds to random $\sigma^{y}_{\alpha,j}$ terms, which have $n^{\,}_x$ and $n^{\,}_z$ both odd. However numerical evaluation of the Fourier coefficients suggests that they fall off with $n^{\,}_{x,z} > 1$ as $e^{-n^2}$ or faster; additionally, the varying sign with $n^{\,}_{x,z}$ leads to further suppression upon summation. Note that $\sigma^{y}_{A,j} \sigma^{y}_{B,j}$ terms commute with both pulses, and while $\sigma^{y}_{B,j} \sigma^{y}_{A,j+1}$ terms do not commute with the generators of the pulses, they have strictly even $n^{\,}_x$ and $n^{\,}_z$ components---because successive Fibonacci numbers cannot both be even, these terms will not have vanishing denominators. While we do not consider them numerically, terms such as $\sigma^{x}_{A,j} \sigma^{z}_{B,j}$ have the same properties as $\sigma^{y}_{\alpha,j}$ perturbations, in terms of decay of Fourier components and overall magnitude (in fact, these terms transform into one another in part upon changing to the interaction frame of $H^{\,}_0$). Thus, for this model, for sufficiently narrow pulses ($w \lesssim T/10$), we do not expect to see divergences due to small $\vecomega \cdot \vecn$ denominators at finite order in the expansion.

\begin{figure}[t]
\centering
\includegraphics[width = 0.5\columnwidth]{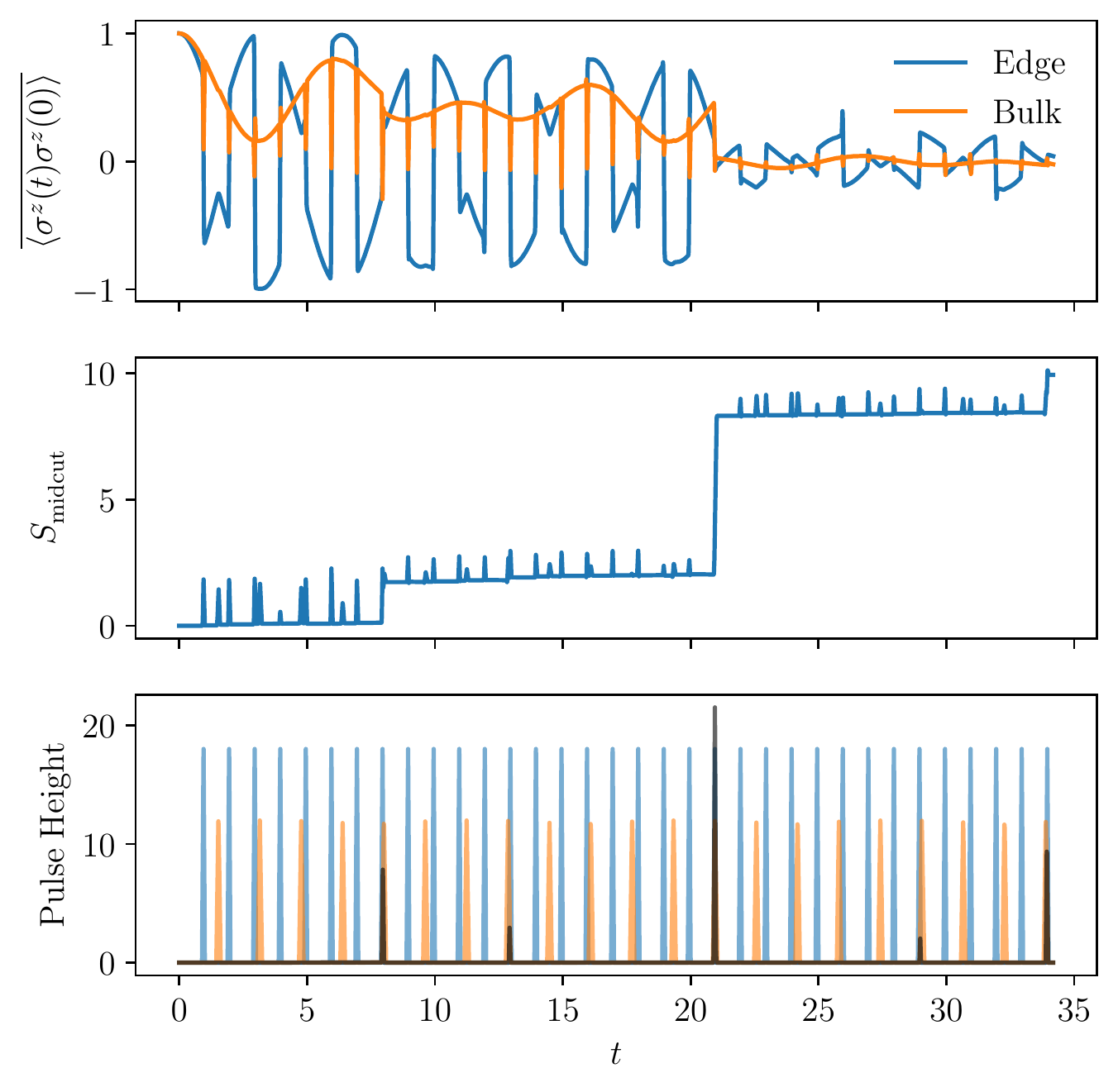} 
\caption{{\bf Entanglement growth from noncommuting pulses} --- TEBD simulations of the dynamics with finite-width, noncommuting $x$ and $z$ pulses. We use a static Hamiltonian consisting of random fields only, with $h^{\,}_x, h^{\,}_y, h^{\,}_z \in [-W, W]$ with $W=0.5$. The drive pulses are triangular, perfect $\pi$ pulses consisting of nearest-neighbor $ZZ$ and $XX$ interactions, with width $T/10$ (where $T$ the period of each pulse). The TEBD parameters are $dt = 0.01$, $\eps=10^{-8}$, and the data were averaged over three disorder realizations.  {\it Top:} Spin correlation functions, {\it Middle:} Half-chain entanglement entropy and {\it Bottom:} Pulse sequence ($x$ pulses in blue, $z$ pulses in orange, and their product in black, indicating when the $x$ and $z$ pulses overlap). The entanglement entropy increases rapidly whenever the noncommuting pulses overlap, signaling that they are incompatible with MBL and lead to thermalization. 
\label{Fig:HeatingNonCommPulses}
}
\end{figure}

\paragraph{Commuting Structure of Pulses} 
We note that, unlike the single-tone Floquet case,  smooth time dependence for multi-tone pulses necessarily requires different pulses to overlap in time. For this reason it is essential that we chose pulse terms in $H^{\,}_0$ that all commute with each other. For example, one could have regrouped the terms in the x,z pulses as single-spin terms: $H^{\prime}_0 = \sum_{\alpha=A,B}\sum_{i=1}^L f^{\,}_{x,z}(t)\sigma^{x,z}(t)$. For a full pulse train (either $x$ or $z$, but not both), this results in the same $\pi$ pulse of $g^{\,}_{x,z}$. 
However, the quasiperiodic sequence of finite-width pulses results in overlap of strong, non-commuting $\sigma^x$ and $\sigma^z$ terms, which we observe (Figure~\ref{Fig:HeatingNonCommPulses}) tend to produce rapid jumps in the entanglement entropy, signaling that these disrupt MBL.

\subsection{Emergent Symmetry Properties} \label{app:emergesymprop}
Intuitively, each term in $D$ consists of terms that are averaged over the $\vec{\theta}$-torus to have net frequency $0$. These terms are ``twirled" over the twisted time translations, $\{ g^{\,}_\mu \}$. One can confirm explicitly that $D$ commutes with the emergent symmetries, $g^{\,}_{x,z}$ \eqref{eq:Q} through analysis of the Fourier transformed quantities, $H^{\prime}_{\vecn}$. Since the pulses commute, we may consider a corresponding integral over one of the $\theta^{\,}_{\nu}$ directions, i.e.,
\be \label{eq:Hnstep1} 
	\int\limits_{0}^{4\pi} \frac{d \theta^{\,}_{\nu}}{4\pi} \, e^{- i n^{\,}_{\nu} \theta^{\,}_{\nu} /2} \, U^{\dagger}_{\nu} \( \theta^{\,}_{\nu} \) \, H^{\prime}_{{\rm S}} \( \theta^{\,}_{\nu}, \theta^{\,}_{\overline{\nu}} \) \, U^{\,}_{\nu} \( \theta^{\,}_{\nu} \),
\ee
where an integral of the above form over both $\theta^{\,}_{x,z}$ defines $ H^{\prime}_{n^{\,}_x, n^{\,}_z}$. 

Defining $\phi^{\,}_{\nu} = \theta^{\,}_{\nu} -2\pi$, we note that $H^{\prime}_{{\rm S}}$ has the same periodicity as $H^{\,}_0$, i.e. $H^{\prime}_{{\rm S}} \( \theta^{\,}_{\nu} + 2\pi \) = H^{\prime}_{{\rm S}} \( \theta^{\,}_{\nu} \)$, and that $U^{\,}_{\nu} \( \theta^{\,}_{\nu} + 2 \pi \) = U^{\,}_{\nu} \( \theta^{\,}_{\nu}  \) \, g^{\,}_{\nu}$. Hence, Eq.~\ref{eq:Hnstep1} can be rewritten as
\begin{gather} \label{eq:Hnstep2}
	\int\limits_{-2\pi}^{2\pi} \frac{d \phi^{\,}_{\nu}}{4\pi} \, e^{i \pi n^{\,}_{\nu} } \, e^{- i n^{\,}_{\nu} \phi^{\,}_{\nu}/2} \, 
	 g^{\,}_{\nu} \, U^{\dagger}_{\nu} \( \phi^{\,}_{\nu} \) \, H^{\prime}_{{\rm S}} \( \phi^{\,}_{\nu}, \theta^{\,}_{\overline{\nu}} \) \, U^{\,}_{\nu} \( \phi^{\,}_{\nu} \) \, g^{\,}_{\nu},
\end{gather}
and extracting the factor of $(-1)^{n^{\,}_{\nu}}$ and the two factors of  $g^{\,}_{\nu}$ from Eq.~\ref{eq:Hnstep2} leads to an integrand that is identical to Eq.~\ref{eq:Hnstep1}. Because both integrands are periodic on the interval of integration, they are equal to one another, i.e. $H^{\prime}_{n^{\,}_{\nu}} (\theta^{\,}_{\overline{\nu}})= e^{i \pi n^{\,}_{\nu} } \, g^{\,}_{\nu} \, H^{\prime}_{n^{\,}_{\nu}} (\theta^{\,}_{\overline{\nu}}) \, g^{\,}_{\nu}$.

Integrating bothEqs.~\ref{eq:Hnstep1} and \ref{eq:Hnstep2} over $\theta^{\,}_{\overline{\nu}}$ (following Eq.~\ref{eq:Hnstep1}), we recover
\be \label{eq:Hnrelation} 
	H^{\prime}_{\vecn} = e^{i \pi n^{\,}_{\nu} } \, g^{\,}_{\nu} \, H^{\prime}_{\vecn} \, g^{\,}_{\nu},
\ee
for either $\nu = x,z$. Fourier terms with $n^{\,}_{\nu}$ \emph{even} then satisfy $H^{\prime}_{\vecn} = g^{\,}_{\nu} \, H^{\prime}_{\vecn} \, g^{\,}_{\nu}$, or equivalently,
\be \label{eq:Hevencomm} 
	\com{H^{\prime}_{\vecn} }{g^{\,}_{\nu}} = 0 \, .~~~
\ee
Importantly, since the Fourier components of the $H^{\prime}_{\vecn}$ factors in $\Dop^{\,}_q$ must sum to $\vecsym{0}$, if terms with \emph{odd} $n^{\,}_{\nu}$ appear, there must be an even number of them, ensuring that 
\be \label{eq:Dcomm} \com{D}{g^{\,}_{\nu}} = 0~,~~\nu = x,z~~.~~\ee

For example, $D^{(2)}_{\,}$ contains sum over $\com{H^{\prime}_{n^{\,}_{\nu}}}{H^{\prime}_{-n^{\,}_{\nu}}}$, with $n^{\,}_{\nu}$ odd; we can then use Eq.~\ref{eq:Hnrelation} to write $\com{ H^{\prime}_{n^{\,}_{\nu}} }{H^{\prime}_{-n^{\,}_{\nu}}}$ as $\com{(-1)^{n^{\,}_{\nu}} \, g^{\,}_{\nu} \, H^{\prime}_{n^{\,}_{\nu}} \, g^{\,}_{\nu}}{ (-1)^{-n^{\,}_{\nu}} \, g^{\,}_{\nu} \, H^{\prime}_{-n^{\,}_{\nu}} \, g^{\,}_{\nu}}$. The factors of $(-1)^{n^{\,}_{\nu}}$ for any $D^{(q)}_{\,}$ can be written as $(-1)^{\sum_{j=1}^{q} n^{\,}_{\nu,j}} \equiv 1$ as a defining property of $D^{(q)}_{\,}$. Since $g^{2}_{\nu} = \ident$, all internal $g^{\,}_{\nu}$ terms cancel, and $\com{\com{H^{\prime}_{n^{\,}_{1,\nu}}}{\dots}}{H^{\prime}_{n^{\,}_{q,\nu}}} = g^{\,}_{\nu} \, \com{\com{H^{\prime}_{n^{\,}_{1,\nu}}}{\dots}}{H^{\prime}_{n^{\,}_{q,\nu}}} \, g^{\,}_{\nu}$, and thus $\com{D^{(q)}_{\,}}{g^{\,}_{\nu}} = 0$

A similar argument can be used to show that $P$ obeys twisted time translation symmetries. In particular, we note that $P = e^{- i \Gamma}$, where each term in $\Gamma^{(q)} ( \vecth )$ can be written in the form
\be \Gamma^{(q)} ( \vecth ) \sim \sum\limits_{\vecn \in \Z \neq 0 } e^{i \vecn \cdot \vecth / 2} \dots ~,~~ \label{eq:Gammaqgen}\ee
where the $\dots$ consist of $q$ denominators involving $\vecomega$ and importantly, nested commutators involving $q$ copies of $H^{\prime}_{\vecm^{\,}_j}$, i.e.,
\be \notag \com{\com{H^{\prime}_{\vecm^{\,}_{1}}}{\dots}}{H^{\prime}_{\vecm^{\,}_{q}}}~,~~\ee
with $\vecn = \sum_{j=1}^{q} \vecm^{\,}_j$. In this case, we are interested in $\Gamma^{(q)} ( \vecth + 2 \pi \vecsym{e}^{\,}_{\nu} )$, which compared to $\Gamma^{(q)} (\vecth)$ imbues the summand in Eq.~\ref{eq:Gammaqgen} with a factor of $\( -1 \)^{n^{\,}_{\nu} }  = \prod_{j=1}^{q} \( -1 \)^{m^{\,}_{j,\nu} } $. 

Just as for $D^{(q)}$, we use the fact that $\com{\com{ \( -1 \)^{m^{\,}_{1,\nu}}  \, H^{\prime}_{m^{\,}_{1,\nu}}}{\dots}}{  \( -1 \)^{m^{\,}_{q,\nu}} \, H^{\prime}_{m^{\,}_{q,\nu}}}$ is equivalent to  $g^{\,}_{\nu} \, \com{\com{H^{\prime}_{m^{\,}_{1,\nu}}}{\dots}}{H^{\prime}_{m^{\,}_{q,\nu}}} \, g^{\,}_{\nu}$ by Eq.~\ref{eq:Hnrelation}, which means that 
\be \Gamma^{(q)} ( \vecth + 2 \pi \vecsym{e}^{\,}_{\nu} ) = g^{\,}_{\nu} \,  \Gamma^{(q)} ( \vecth ) \,  g^{\,}_{\nu} ~~.~~\ee
since $P = \exp ( -i \sum_{q=1} \Gamma^{(q)} )$, we have 
\be P ( \vecth + 2 \pi \vecsym{e}^{\,}_{\nu} ) = g^{\,}_{\nu} \, P ( \vecth ) \, g^{\,}_{\nu} ~~,~~\ee
at any given order (i.e., $P$ obeys twisted time translation symmetries). 

However, because $U^{\,}_0 (\vecth + 2 \pi \vecsym{e}^{\,}_{\nu} ) = U^{\,}_0 (\vecth) g^{\,}_{\nu}$, we find that
\be Q (\vecth + 2 \pi \vecsym{e}^{\,}_{\nu} ) = W \, U^{\vpd}_0 (\vecth) g^{2}_{\nu}  P ( \vecth )  g^{2}_{\nu} U^{\dagger}_0 (\vecth) = Q (\vecth),~~~~~\ee
i.e., $Q$ has the time translation properties of the original Hamiltonian.

\section{Absence of Floquet EDSPTs \label{app:FloqCD}}
Here we provide a simple argument that any gapped phase without an anomalous edge-states is continuously (without a gap closing) connected to a trivial insulator (product-state ground-state), and similarly any MBL system (including periodic and quasiperiodically driven ones) without anomalous edge-states is continuously connected to a trivial MBL system (with all eigenstates being product states). We then show how this mechanism can be used to trivialize an attempted Floquet EDSPT construction, whose generalization to general group-cohomology classes suggests that Floquet EDSPTs are impossible and that EDSPTs are special to quasiperiodically driven settings.

\subsection{SPTs without anomalous edges can be trivialized}
Consider a gapped (or MBL) system that lacks anomalous edge states, i.e. for which it is possible to continuously deform the edge to a trivial product state with edge-local perturbations or counter-drives. Denote the correlation length or localization length of the initial system by $\xi$. Then, consider selecting a regular array of finite size blocks of linear-dimension $\ell\sim \xi$, where each block is separated from the others by distance $x\gg \xi$, and continuously interpolating the local Hamiltonian within those blocks to a trivial one. Since the blocks have fixed, finite-size, and are well separated, this does not result in a phase transition (for sufficiently large $x$). This results in a ``Swiss-cheese" like arrangement of holes, filled with trivial unentangled matter. By assumption we can trivialize the interface of each hole since there is no anomaly obstruction. By repeating this process, we can trivialize more and more parts of the system, until eventually (in $\mathcal{O}(x/\ell)^d$ steps, where $d$ is the spatial dimensionality), the entire system is trivial. This process provides a continuous path to trivialize the initial system, while maintaining a (mobility) gap throughout, i.e. proves that the initial system was in a trivial phase. In contrast, with anomalous edge states, this procedure produces a finite density of gapless interface states that will percolate through the sample at some step in the process, resulting in a phase transition. Note that, for intrinsic topological orders, this procedure would result in a very high-genus surface with extensive ground-state degeneracy, and would also fail even in the absence of gapless interfaces.

This argument shows that an edge-anomaly is essential for the stability of a non-trivial invertible topological phase. As an immediate corollary, to demonstrate that a phase is trivial, it is sufficient to show that its edge can be deformed to a trivial one by local interactions. In the next section, we will use this strategy to analytically show that a Floquet analog of our construction in the main-text fails to produce a non-trivial EDSPT.

\subsection{Reminder: Levin-Gu Phase}
In a pioneering work~\cite{levin2012braiding}, Levin and Gu constructed a model of a 2d bosonic SPT protected by a single $\Z_2$ symmetry (henceforth referred to as the Levin-Gu (LG) model). The LG model consists of spins-1/2 on a triangular lattice, with Hamiltonian:
\begin{align}
	H &= -\sum_i \lambda_i \tilde{\sigma}^x_i, 
	\nonumber\\
	\tilde\sigma^x_i &= \prod_{\<kl\>\in \raisebox{-0.02in}{\Large \hexagon} \hspace{-0.09in}i} i^{\frac12(1-\sigma^z_k\sigma^z_l)} 
	\sigma^x_i
	\label{eq:HLG}
\end{align}
where the product in the second line ranges over the links on the hexagon of nearest-neighbors to site $i$, and we have allowed for spatially dependent coupling constants $\lambda_i$ to permit MBL-stabilization of excited state SPT order. The argument of the phase-factor exponent counts the number of domain walls (DWs) on the perimeter of the hexagonal plaquette surrounding $i$, which is necessarily even, we can write the phase as: $(-1)^{\# \text{DWs}/2}$.

This model has an ordinary microscopic $\Z_2$ symmetry generated by $g=\prod_i\sigma^x_i$. The effect of the phase factors in the second line of Eq.~\ref{eq:HLG} can be understood by gauging this symmetry, in which case $\Z_2$-symmetry fluxes become Abelian anyons (semions), whose Abelian braiding statistics is manifest in the fusion rules for the intersection of $\Z_2$-DWs with the sample boundary in the original, ungauged SPT model.

For sites near an open boundary, \label{eq:HLG} is ambiguous due to incomplete hexagonal plaquettes. Following \cite{levin2012braiding}, one can define $\tilde\sigma^x_i$ for boundary sites by adopting the convention that all sites $j,k$ lying outside the system are taken to have non-dynamical ``ghost" spins that are pointing up in the $z$-direction. This choice clearly hides the $\Z_2$ symmetry, and will result in a non-trivial symmetry-transformation of boundary degrees of freedom:
\begin{align}
	g\, \tilde\sigma^x_{i\in \text{bdy}} \, g = -\sigma^z_{i+1}\tilde\sigma^x_i\sigma^z_{i-1}
\end{align}
where we have ordered the indices, $i, i\pm 1$ along the boundary (the choice of orientation is not important in for this $\Z_2$ example). Note also, that $\sigma^z$ has the same commutation relations, $\{\sigma^z_i,\tilde\sigma^x_i\}=0$, with $\tilde\sigma^x$ as with $\sigma^x$. Connoisseurs of SPT will recognize this transformation as implementing a duality transformation between the trivial paramagnetic (PM) terms $\tilde\sigma^x$ and the $1d$ cluster-state (CS) terms. 

The same transformation can be implemented by a unitary acting only in a finite strip near the edge:
\begin{align}
	&g\, \tilde\sigma^x_{i\in \text{bdy}} \, g = V\tilde\sigma^x_iV^\dagger \nonumber\\
	&V = \prod_{i \in \text{bdy}}e^{i\frac\pi4 \sigma^z_i(1-\sigma^z_{i-1}\sigma^z_{i+1})}.
\end{align}
$V$ acts only on unit cells that overlap the system boundary, and is trivial in the bulk. For future use, note that $gVg=V^\dagger$.

\subsection{(Failed) Prototype of a a Floquet EDSPT}
We attempt to promote the static LG-model to a Floquet model, where the symmetry is dynamically enforced by $\pi$-pulses of $g$. Consider a stroboscopic Floquet lattice model defined on an open domain $\Sigma$, whose Floquet operator (time-evolution for one period, $T$) is:
\begin{align}
	U(T) = ge^{-i(H_\Sigma + H_{\d\Sigma})}
\end{align}
where we have separated Eq.~\ref{eq:HLG} into bulk: $H_\Sigma = \sum_{i\in \text{Int}(\Sigma)}\lambda_i\tilde\sigma^x_i$, 
and boundary: 
$H_{\d\Sigma} =
 \sum_{i\in \d\Sigma} \lambda_i\tilde{\sigma}^x_i
$ 
where $\text{Int}(\Sigma)$ and $\d\Sigma$ respectively denote the interior and boundary of $\Sigma$.

By inspection, one can see that this particularly boundary termination yields a non-trivial (thermal or spontaneous dynamical symmetry-breaking) boundary by considering evolution for two periods:
\begin{align}
	U(2T) = e^{-i 2H_\Sigma}
	e^{ \sum_{i\in \d\Sigma} \lambda_i\tilde{\sigma}^x_i}
	e^{ \sum_{i\in \d\Sigma} \lambda_i\sigma^z_{i-1}\tilde{\sigma}^x_i\sigma^z_{i+1}}
\end{align}
the latter two terms are related by a generalized Kramers-Wannier duality that exchanges PM and SPT phases, such that the resulting boundary theory is self-dual. As discussed in \cite{potter2017dynamically}, and building on results from~\cite{friedman2018localization,prakash2017eigenstate}, this self-duality produces a local symmetry-enforced degeneracy on the boundary, which prevents the boundary from obtaining a trivial, symmetric MBL state. 

So far, we have considered a fine-tuned version of this model with a microscopic $\Z_2$ symmetry. Now consider breaking this symmetry by arbitrary, but \emph{weak} perturbations. The HF-expansion outlined above, implies that, up to some prethermal time-scale, the above-outlined physics survives, with an emergent dynamical symmetry enforced by the $g$-pulses. In the next section, we consider \emph{strong} deformation of the edge drive (beyond the purview of the HF-expansion), which we can analytically show destroys the edge model at a special solvable point.

\subsection{Absence of edge-anomaly in a Floquet Levin-Gu phase without symmetry}
To trivialize the edge, we consider applying an extra step of stroboscopic evolution which undoes the duality transformation on the edge spins implemented by $g$:
\begin{align}
	U'(T) = Vge^{-i(H_\Sigma + H_{\d\Sigma})}.
\end{align}
Notice, that only the edge has been modified, and the bulk remains the same.

To analyze the spectrum of this model, it is again convenient to consider the two-period evolution operator:
\begin{align}
	U'(2T) 	&= e^{-i 2H_\Sigma}Vge^{-iH_{\d\Sigma}}Vge^{-i H_{\d\Sigma}}
	\nonumber\\
			&= e^{-i 2H_\Sigma}Vge^{-iH_{\d\Sigma}}gV^\dagger e^{-i H_{\d\Sigma}}
	\nonumber\\
			& = e^{-i 2(H_\Sigma+H_{\d\Sigma})}
\end{align}
where in the second line we have inserted $g^2=1$, and used that $gVg = V^\dagger$, and in the last line we have noted that conjugation by $g$ and $V$ have compensating effects on $\tilde\sigma^x_i$ for boundary spins: $Vg\tilde\sigma^x_{i\in \d\Sigma} gV^\dagger =\tilde\sigma^x_{i\in \d\Sigma}$. 

Examining the final line, we see that the resulting edge terminates with a trivial, MBL paramagnetic phase, which, in the absence of any microscopic symmetry can be disentangled with a finite-depth local unitary acting only the boundary. Together with the above arguments of the previous sections, this demonstrates that the $g$-pulses are insufficient to dynamically enforce a $\Z_2$ symmetry that protects anomalous edge behavior, and that the putative Floquet EDSPT is, in fact, trivial.

Compared to the $1d$ quasiperiodic example described in the main text, this Floquet example has the crucial distinction that the counter-drive can be applied as a separate stroboscopic step, without requiring non-smooth $\delta$-function pulses (e.g. the extra stroboscopic step can be applied with a smooth bump function time-profile which has stretched-exponentially decaying frequency content), and without resulting in overlap of non-commuting pulses (which we saw, in the quasiperiodic case, led to thermalization).

While we have worked out the case explicitly for the Levin-Gu model, this model is indicative of the structure of other exactly solvable models of phases classified by group-cohomology~\cite{chen2013symmetry}, and a similar construction works more generally to trivialize putative Floquet SPTs in all cohomology classes. Since, beyond-cohomology classes do not permit MBL due to the presence of chiral surface modes~\cite{potter2015protection}, this exhausts the possibilities for bosonic SPTs, and shows that Floquet EDSPTs are not possible for interacting bosonic systems.

\subsection{Contrasting Periodic and Quasiperiodic Drives}
For static or Floquet systems, we argued above that the ability to turn on edge interactions to  trivially gap-out or localize the edge of a system led to a route to trivializing the bulk without a phase transition, by punching out a sequence of non-overlapping trivial holes, and healing the interface with the edge-trivializing procedure. Importantly, this mechanism enabled the bulk to be trivialized even when the boundary trivialization procedure necessarily passes through a boundary phase transition en route to the trivial edge. Namely, at any stage in the process, the bulk ``swiss-cheese'' version of this procedure only ever modifies on finite-size $0d$ chunks of the system. In static and Floquet systems, finite-size $0d$ systems cannot undergo criticality, and hence the gapless/delocalized critical modes that might be encountered upon trivializing an infinite boundary, are avoided when trivializing the bulk.

In contrast, in the quasiperiodic system analyzed in the main text, we find evidence that turning on the edge counter drive induces a $0d$ quasiperiodic-to-chaotic dynamical transition. Let us assume for the moment that our numerical evidence reflects a true $0d$ phase transition rather than a finite-size artifact. Then would imply that attempting to trivialize the bulk via this mechanism would require introducing chaotic spins that produce a bulk delocalization transition, so that the bulk-trivialization procedure fails to smoothly deform the quasiperiodic EDSPT to a trivial phase without encountering a bulk phase transition. Ironically, while this mechanism highlights the relative fragility of localization in quasiperiodically driven signatures, it would actually protect a finer distinction among quasiperiodic dynamical phases of matter!

We close by noting, that regardless of whether there is a true boundary phase transition, the HF-expansion above shows that weak perturbations from any solvable drive lead to (stretched)-exponentially long lived phenomena, which in a practical sense can be stable to very long time scales that greatly exceed experimental lifetimes, or other more pressing dangers to localization (like inevitable weak-coupling to the environment).

\end{document}